\magnification=1200
\def\tr#1{{\rm tr} #1}
\def\d{\partial}
\def\f#1#2{{\textstyle{#1\over #2}}}
\def\st#1{{}^{(4)}#1}
\def\next{\hfil\break\noindent}
\def\R{{\bf R}}
\font\title=cmbx12
   
\ 
\vskip 3cm
\centerline{\title An introduction to the Einstein-Vlasov system}

\vskip 20pt
\centerline{Alan D. Rendall}
\centerline{Max-Planck-Institut f\"ur Gravitationsphysik}
\centerline{Schlaatzweg 1}
\centerline{14473 Potsdam, Germany}

\vskip 30pt
\centerline{\bf Abstract}

\vskip 20pt
These lectures are designed to provide a general introduction to the
Einstein-Vlasov system and to the global Cauchy problem for these
equations. To start with some general facts are collected and a
local existence theorem for the Cauchy problem stated. Next the case
of spherically symmetric asymptotically flat solutions is examined in
detail. The approach taken, using maximal-isotropic coordinates, is
new. It is shown that if a singularity occurs in the time evolution
of spherically symmetric initial data, the first singularity (as measured
by a maximal time coordinate) occurs at the centre. Then it is shown that
for small initial data the solution exists globally in time and is 
geodesically complete. Finally, the proof of the general local existence
theorem is sketched. This is intended to be an informal introduction to 
some of the ideas which are important in proving such theorems rather than 
a formal proof.

\vfil\break

\noindent
{\bf 1. Introduction}

The Vlasov equation arises in kinetic theory. It gives a statistical
description of a collection of particles. It is distinguished from
other equations of kinetic theory by the fact that there is no direct
interaction between particles. In particular, no collisions are
included in the model. Each particle is acted on only by fields which
are generated collectively by all particles together. The fields which
are taken into account depend on the physical situation being
modelled. In plasma physics, where this equation is very important,
the interaction is electromagnetic and the fields are described either
by the Maxwell equations or, in a quasi-static approximation, by the
Poisson equation [26]. In gravitational physics, which is the subject
of the following, the fields are described by the Einstein equations
or, in the Newtonian approximation, by the Poisson equation. (There is
a sign difference in the Poisson equation in comparison with the
electromagnetic case due to the replacement of a repulsive by an
attractive force.) The best known applications of the Vlasov equation
to self-gravitating systems are to stellar dynamics[3]. It can also be
applied to cosmology.  In the first case the systems considered are
galaxies or parts of galaxies where there is not too much dust or gas
which would require a hydrodynamical treatment. Possible applications
are to globular clusters, elliptical galaxies and the central bulge of
spiral galaxies.  The \lq particles\rq\ in all these cases are
stars. In the cosmological case they are galaxies or even clusters of
galaxies. The fact that they are modelled as particles reflects the
fact that their internal structure is believed to be irrelevant for
the dynamics of the system as a whole.

These lectures are concerned not with the above physical applications 
but with some basic mathematical aspects of the Einstein-Vlasov 
system. First the definition and general mathematical properties of
this system of partial differential equations are discussed and then
the Cauchy problem for this system is formulated. The central theme
in what follows is the global Cauchy problem, where \lq global\rq\ means 
global in time. Up to now global results have only been obtained in
very special cases and one of these, the case of spherically symmetric
asymptotically flat solutions, is discussed in detail. The global
existence results presented depend on having a suitable local existence
result. At first the necessary local existence theorem is stated without
proof and used in obtaining global theorems. The proof of the local
existence theorem is discussed in some detail afterwards. Further
information on kinetic theory in general relativity may be found in [10].

Let $(M,g_{\alpha\beta})$ be a spacetime, i.e. $M$ is a four-dimensional 
manifold and $g_{\alpha\beta}$ is a metric of Lorentz signature $(-,+,+,+)$.
Note that $g_{\alpha\beta}$ denotes a geometric object here and not the
components of the geometric object in a particular coordinate system.
In other words the indices are abstract indices. (See [28], section 2.4
for a discussion of this notation.)
It is always assumed that the metric is time-orientable, i.e. that the 
two halves of the light cone at each point of $M$ can be labelled past
and future in a way which varies continuously from point to point.
With this global direction of time, it is possible to distinguish
between future-pointing and past-pointing timelike vectors. The worldline of 
a particle of non-zero rest mass $m$ is a timelike curve in spacetime. The 
unit future-pointing tangent vector to this curve is the 4-velocity 
$v^\alpha$ of the particle. Its 4-momentum $p^\alpha$ is given by 
$mv^\alpha$. There are different variants of
the Vlasov equation depending on the assumptions made. Here it is 
assumed that all particles have the same mass $m$ but it would also
be possible to allow a continuous range of masses. When all the 
masses are equal, units can be chosen so that $m=1$ and no distinction
need be made between 4-velocity and 4-momentum. There is also the 
possibility of considering massless particles, whose wordlines are
null curves. In the case $m=1$ the possible values of the four-momentum
are precisely all future-pointing unit timelike vectors. These form
a hypersurface $P$ in the tangent bundle $TM$ called the mass shell.
The distribution function $f$, which represents the density of particles
with given spacetime position and four-momentum, is a non-negative
real-valued function on $P$. A basic postulate in general relativity is
that a free particle travels along a geodesic. Consider a future-directed
timelike geodesic parametrized by proper time. Then its tangent vector
at any time is future-pointing unit timelike. Thus this geodesic has a
natural lift to a curve on $P$, by taking its position and tangent
vector together. This defines a flow on $P$. Denote the vector field
which generates this flow by $X$. (This vector field is what is sometimes
called the geodesic spray in the mathematics literature.) The condition
that $f$ represents the distribution of a collection of particles moving
freely in the given spacetime is that it should be constant along the
flow, i.e. that $Xf=0$. This equation is the Vlasov equation, sometimes
also known as the Liouville or collisionless Boltzmann equation.

To get an explicit expression for the Vlasov equation, it is necessary
to introduce local coordinates on the mass shell. In the following 
local coordinates $x^\alpha$ on spacetime are always chosen such that
the hypersurfaces $x^0$=const. are spacelike. (Greek and Roman indices take 
the values $0,1,2,3$ and $1,2,3$ respectively.) Intuitively this means that 
$x^0$, which may also be denoted by $t$, is a time coordinate and that the 
$x^a$ are spatial coordinates. A timelike vector is future-pointing if and 
only if its zero component in a coordinate system of this type is positive.
It is not assumed that the vector $\d /\d x^0$ is timelike. One way of 
defining local coordinates on $P$ is to take the spacetime coordinates 
$x^\alpha$ together with the spatial components $p^a$ of the four-momentum 
in these coordinates. Then the explicit form of the Vlasov equation is:
$$\d f/\d t+(p^a/p^0)\d f/\d x^a-\Gamma^a_{\beta\gamma}p^\beta p^\gamma
\d f/\d p^a=0\eqno(1.1)$$
where $\Gamma^\alpha_{\beta\gamma}$ are the Christoffel symbols
associated to the metric $g_{\alpha\beta}$.  Here it is understood
that $p^0$ is to be expressed in terms of $p^a$ and the metric using
the relation $g_{\alpha\beta}p^\alpha p^\beta=-1$.  An alternative way
of coordinatizing the mass shell which is often useful is to use the
components of the four-momentum in an orthormal frame, which has a
priori nothing to do with the frame defined by the coordinates. It
should be chosen so that the first vector is future-pointing timelike.
The explicit form of the Vlasov equation in these coordinates is
similar to (1.1), with the Christoffel symbols replaced by the Ricci
rotation coefficients $\gamma^\mu_{\nu\rho}$ of the frame. Explicitly, it 
is given
by:
$$\d f/\d t+(v^i/v^0)e_i^a\d f/\d x^a-\gamma^i_{\mu\nu}v^\mu v^\nu
\d f/\d v^i=0\eqno(1.2)$$
The convention is used that frame indices are denoted by letters from
the middle of the alphabet, while coordinate indices are taken from the
beginning of the alphabet. The components of the frame vectors are denoted
by $e^\alpha_\mu$ and $p^\alpha$ and $v^\mu$ are related by the equation
$p^\alpha=e^\alpha_\mu v^\mu$. Since the frame is orthonormal 
$v^0=\sqrt{1+\delta^{ij}v_iv_j}$, where $\delta^{ij}$ is the Kronecker
delta. 

The Vlasov equation can be coupled to the Einstein equations as follows,
giving rise to the Einstein-Vlasov system. The unknowns are a 4-manifold
$M$, a (time orientable) Lorentz metric $g_{\alpha\beta}$ on $M$ and a 
non-negative real-valued function $f$ on the mass shell defined by 
$g_{\alpha\beta}$. The field equations consist of the Vlasov equation 
defined by the metric $g_{\alpha\beta}$ for
$f$ and the Einstein equation $G_{\alpha\beta}=8\pi T_{\alpha\beta}$.
(Units are chosen here so that the speed of light and the gravitational
constant both have the numerical value unity.)
To obtain a complete system of equations it remains to define $T_{\alpha
\beta}$ in terms of $f$ and $g_{\alpha\beta}$. It is defined as an integral 
over the part of the mass shell over a given spacetime point with respect to 
a measure which will now be defined. The metric at a given point of 
spacetime defines in a tautological way a metric on the tangent space
at that point. The part of the mass shell over that point is a 
submanifold of the tangent space and as such has an induced metric, 
which is Riemannian. The associated measure is the one which we are
seeking. It is evidently invariant under Lorentz transformations of
the tangent space, a fact which may be used to simplify computations in 
concrete situations. In the coordinates $(x^\alpha,p^a)$ on $P$ the
explicit form of the energy-momentum tensor is:
$$T_{\alpha\beta}=-\int fp_\alpha p_\beta |g|^{1/2}/p_0 dp^1 dp^2 dp^3
\eqno(1.3)$$
A simple computation in normal coordinates based at a given point shows
that $T_{\alpha\beta}$ defined by (1.3) is divergence-free, independently
of the Einstein equations being satisfied. This is of course a necessary
compatibility condition in order for the Einstein-Vlasov system to be
a reasonable set of equations. Another important quantity is the
particle current density, defined by:
$$N^\alpha=-\int fp^\alpha |g|^{1/2}/p_0 dp^1 dp^2 dp^3\eqno(1.4)$$ 
A computation in normal coordinates shows that $\nabla_\alpha N^\alpha=0$.
This equation is an expression of the conservation of the number of
particles. There are some inequalities which follow immediately from
the definitions (1.3) and (1.4). Firstly $N_\alpha V^\alpha\le 0$ for any
future-pointing timelike or null vector $V^\alpha$, with equality only if 
$f=0$ at the given point. Hence unless there are no particles at some point, 
the vector $N^\alpha$ is future-pointing timelike. Next, if $V^\alpha$ and
$W^\alpha$ are any two future-pointing timelike vectors then
$T_{\alpha\beta}V^\alpha W^\beta\ge 0$. This is the dominant energy
condition ([12], p. 91). Finally, if $X^\alpha$ is a spacelike vector then
$T_{\alpha\beta}X^\alpha X^\beta\ge 0$. This is the non-negative
pressures condition. This condition, the dominant energy condition and
the Einstein equations together imply that the Ricci tensor satisfies the
inequality $R_{\alpha\beta}V^\alpha V^\beta\ge 0$ for any timelike vector 
$V^\alpha$. The last inequality is called the strong energy condition.
These inequalities constitute one of the reasons
which mean that the Vlasov equation defines a well-behaved matter model
in general relativity. However this is not the only reason. A perfect
fluid with a reasonable equation of state or matter described by the 
Boltzmann equation also have energy-momentum tensors which satisfy these 
inequalities.

The Vlasov equation in a fixed spacetime is a linear hyperbolic equation
for a scalar function and hence solving it is equivalent to solving
the equations for its characteristics. In coordinate components 
these are:
$$\eqalign{
dX^a/ds&=P^a   \cr
dP^a/ds&=-\Gamma^a_{\beta\gamma}P^\beta P^\gamma}\eqno(1.5)$$
Let $X^a(s,x^\alpha,p^a)$, $P^a(s,x^\alpha,p^a)$ be the unique 
solution of (1.5) with initial conditions $X^a(t,x^\alpha,p^a)=x^a$
and $P^a(t,x^\alpha,p^a)=p^a$. Then the solution of the Vlasov equation can
be written as:
$$f(x^\alpha,p^a)=f_0(X^a(0,x^\alpha,p^a),P^a(0,x^\alpha,p^a))\eqno(1.6)$$
where $f_0$ is the restriction of $f$ to the hypersurface $t=0$. This
function $f_0$ serves as initial datum for the Vlasov equation.
It follows immediately from this that if $f_0$ is bounded by some constant
$C$, the same is true of $f$. This obvious but important property of the 
solutions of the Vlasov equation is used frequently without comment in the 
study of this equation.

The above calculations involving $T_{\alpha\beta}$ and $N^\alpha$ were only
formal. In order that they have a precise meaning it is necessary to
impose some fall-off in the momentum variables on $f$ so that the integrals
occurring exist. The simplest condition to impose is that $f$ has compact
support for each fixed $t$. This property holds if the initial datum $f_0$ 
has compact support and if each hypersurface $t=t_0$ is a Cauchy hypersurface.
For by the definition of a Cauchy hypersurface, each timelike curve which 
starts at $t=0$ hits the hypersurface $t=t_0$ at a unique point. Hence the
geodesic flow defines a continuous mapping from the part of the mass shell
over the initial hypersurface $t=0$ to the part over the hypersurface $t=t_0$.
The support of $f(t_0)$, the restriction of $f$ to the hypersurface $t=t_0$
is the image of the support of $f_0$ under this continuous mapping and so
is compact. Let $P(t)$ be the supremum of the values of $|p^a|$ attained on
the support of $f(t)$. It turns out that in many cases controlling the
solution of the Vlasov equation coupled to some field equation in the
case of compactly supported initial data for the distribution function can be 
reduced to obtaining a bound for $P(t)$. An example of this is given below.

The data in the Cauchy problem for the Einstein equations coupled to any
matter source consist of the induced metric $g_{ab}$ on the initial 
hypersurface, the second fundamental form $k_{ab}$ of this hypersurface and 
some matter data. In fact these objects should be thought of as objects on
an abstract 3-dimensional manifold $S$. Thus the data consist of a 
Riemannian metric $g_{ab}$, a symmetric tensor $k_{ab}$ and appropriate
matter data, all defined intrinsically on $S$. The nature of the initial
data for the matter will now be examined in the case of the Einstein-Vlasov
system. It is not quite obvious what to do, since the distribution function
$f$ is defined on the mass shell and so the obvious choice of initial data,
namely the restriction of $f$ to the initial hypersurface, is not 
appropriate. For it is defined on the part of the mass shell over the
initial hypersurface and this is not intrinsic to $S$. This difficulty
can be overcome as follows. Let $\phi$ be the mapping which sends a 
point of the mass shell over the initial hypersurface to its orthogonal
projection onto the tangent space to the initial hypersurface. The map
$\phi$ is a diffeomorphism. The abstract initial datum $f_0$ for $f$ is
taken to be a function on the tangent bundle of $S$. The initial
condition imposed is that the restriction of $f$ to the part of the
mass shell over the initial hypersurface should be equal to $f_0$ composed
with $\phi$. An initial data set for the Einstein equations must satisfy
the constraints and in order that the definition of an abstract initial
data set for the Einstein equations be adequate it is necessary that the
constraints be expressible purely in terms of the abstract initial data.
The constraint equations are:
$$\eqalignno{R-k_{ab}k^{ab}+(\tr k)^2&=16\pi\rho&(1.7)       \cr
\nabla_a k^a_b-\nabla_b(\tr k)&=8\pi j_b&(1.8)}$$
Here $R$ denotes the scalar curvature of the metric $g_{ab}$. If $n^\alpha$
denotes the future-pointing unit normal vector to the initial hypersurface
and $h^{\alpha\beta}=g^{\alpha\beta}+n^\alpha n^\beta$ is the orthogonal
projection onto the tangent space to the initial hypersurface then 
$\rho=T_{\alpha\beta}n^\alpha n^\beta$ and $j^\alpha=-h^{\alpha\beta}
T_{\beta\gamma}n^\gamma$. The vector $j^\alpha$ satisfies 
$j^\alpha n_\alpha=0$ and so can be naturally identified with a vector
intrinsic to the initial hypersurface, denoted here by $j^a$. What needs
to be done is to express $\rho$ and $j_a$ in terms of the intrinsic initial
data. They are given by the following expressions:
$$\eqalignno{
\rho&=\int f_0(p^a)p^a p_a/(1+p^a p_a)^{1/2}({}^{(3)}g)^{1/2} 
dp^1 dp^2 dp^3
&(1.9)                         \cr
j_a&=\int f_0(p^a)p_a ({}^{(3)}g)^{1/2} dp^1 dp^2 dp^3&(1.10)}$$
If a three-dimensional manifold on which an initial data set for the
Einstein-Vlasov system is defined is mapped into a spacetime by an
embedding $\psi$ then the embedding is said to induce the given initial
data on $S$ if the induced metric and second fundamental form of $\psi(S)$ 
coincide with the results of transporting $g_{ab}$ and $k_{ab}$ with $\psi$ 
and the relation $f=f_0\circ\phi$ holds, as above. A form of the local 
existence and uniqueness theorem can now be stated. This will only be done 
for the case of smooth (i.e. infinitely differentiable) initial data
although versions of the theorem exist for data of finite differentiability.

\vskip 10pt\noindent
{\bf Theorem 1.1} Let $S$ be a 3-dimensional manifold, $g_{ab}$ a smooth
Riemannian metric on $S$, $k_{ab}$ a smooth symmetric tensor on $S$
and $f_0$ a smooth non-negative function of compact support on the
tangent bundle $TS$ of $S$. Suppose further that these objects satisfy
the constraint equations (1.7)-(1.8). Then there exists a smooth spacetime 
$(M,g_{\alpha\beta})$, a smooth distribution function $f$ on the 
mass shell of this spacetime and a smooth embedding $\psi$ of $S$ into $M$ 
which induces the given initial data on $S$ such that $g_{\alpha\beta}$
and $f$ satisfy the Einstein-Vlasov system and $\psi(S)$ is a Cauchy 
hypersurface. Moreover, given any other spacetime $(M',g'_{\alpha\beta})$, 
distribution function $f'$ and embedding $\psi'$ satisfying these conditions, 
there exists a diffeomorphism $\chi$ from an open neighbourhood of 
$\psi(S)$ in $M$ to an open neighbourhood of $\psi'(S)$ in $M'$ which 
satisfies $\chi\circ\psi=\psi'$ and carries $g_{\alpha\beta}$ and $f$ to 
$g'_{\alpha\beta}$ and $f'$ respectively.

\vskip 10pt\noindent
The formal statement of this theorem is rather complicated, but its 
essential meaning is as follows. Given an initial data set (satisfying
the constraints) there exists a corresponding solution of the Einstein-Vlasov 
system and this solution is locally unique up to diffeomorphism. There also 
exists a global uniqueness statement which uses the notion of the maximal 
Cauchy development of an initial data set, but this is not required in 
the following.  The first proof of a theorem of this kind for the 
Einstein-Vlasov system is due to Choquet-Bruhat [7].

In the following we are mainly concerned with asymptotically flat
spacetimes. These are the spacetimes which are appropriate for describing
isolated systems in general relativity. It is assumed that these
spacetimes admit a Cauchy hypersurface with topology $\R^3$, although more
general cases could also be considered. The smooth data set $(g_{ab},
k_{ab}, f_0)$ on $\R^3$ is said to be asymptotically
flat if there exist global coordinates $x^a$ such that as $|x|$ tends
to infinity the components $g_{ab}$ in these coordinates tend to 
$\delta_{ab}$, the components $k_{ab}$ tend to zero, $f_0$ has compact
support and certain norms are finite. These are weighted Sobolev norms.
If $u$ is a smooth function on $\R^3$ define
$$\|u\|_{H^s_\delta}=\left[\sum_{i=0}^s \int (1+|x|^2)^{(\delta+i)}
|D^i u|^2 dx\right]^{1/2}\eqno(1.11)$$ 
where $|D^i u|$ denotes the maximum modulus of any partial derivative of
order $i$ of the function $u$. If the quantity (1.11) is finite then $u$ is
said to belong to the weighted Sobolev space $H^s_\delta$. Assume for 
asymptotic flatness that $g_{ab}-\delta_{ab}\in H^s_\delta$ and 
that $k_{ab}\in H^{s-1}_{\delta+1}$ for $s$ sufficiently large 
and $-3/2<\delta<-1/2$. If a spacetime is asymptotically flat then
Theorem 1.1 can be sharpened to say that there exists a local solution 
corresponding to the given initial data and coordinates defined
for that solution such that the solution exists on $\R^3\times [0,T)$
for some $T>0$ and the data induced on the hypersurfaces $t=$const.
satisfy the same type of asymptotic flatness conditions as the initial data 
on the hypersurface $t=0$. More precisely, there exist coordinates
so that the data induced by the solution on each hypersurface of constant
time belongs to the same weighted Sobolev space as the initial data.
Moreover the restrictions of the functions $g_{00}+1$ and $g_{0a}$ of the
metric components to any hypersurface of constant time belong to weighted 
Sobolev spaces. A proof of this is sketched in Section 4. 

The local existence theorem can be supplemented by a statement that the
solution depends continuously on the data. This is not stated in full
generality here; only some statements for asymptotically flat data are 
given. Given a family of initial data which is bounded in a weighted Sobolev 
space with $s$ sufficiently large and is such that the metric is uniformly
positive definite, the coordinates above can be chosen so that the solutions 
corresponding to all data in the family exist on the same time interval 
$[0,T)$, the weighted Sobolev norm of the data induced by these solutions on 
a hypersurface of constant time is uniformly bounded and the induced metric
on one of these hypersurfaces is uniformly positive definite. Moreover, the 
restrictions of $g_{00}+1$ and $g_{0a}$ to each hypersurface of constant time 
are uniformly bounded in a weighted Sobolev space for all data in the family 
and $g_{00}$ is uniformly bounded away from zero. See Section 4.

Suppose now that an asymptotically flat initial data set admits a group
$G$ of symmetries, i.e. that a Lie group $G$ acts on the manifold $S$
in such a way that $g_{ab}$, $k_{ab}$ and $f_0$ are preserved. Then the
spacetime in Theorem 1.1 can be chosen so that it admits $G$ as a symmetry
group. More precisely, there exists an action of $G$ on $M$ by isometries
which preserves $f$ and which restricts to the original action on $S$.
To prove this, consider the geodesic $\gamma (p)$ through a point $p\in S$ 
orthogonal to $S$ in a spacetime with the given initial data. Let $t(p)$ 
denote the largest number such that, when $\gamma (p)$ is parametrized by
proper time, with $p$ corresponding to the parameter value zero, this 
geodesic is defined on the interval $(-t(p),t(p))$. By what has been said
above, the spacetime can be chosen so that there exists a number $T>0$
which is smaller than $t(p)$ for each $p\in S$. Since the spacetime is
globally hyperbolic, each point can be joined to the initial hypersurface
by a timelike geodesic of maximal length. If $T$ is chosen sufficiently small
then the geodesic is unique. A new spacetime can be defined as the open subset
of the original spacetime where this distance is less than $T$. It will now be
shown that the action of $G$ on $S$ extends to an action on this new 
spacetime, whose underlying manifold will be denoted by $M$. Given a point 
$q\in M$, let $p$ be the point where the unique geodesic $\gamma$ of maximal 
length from $q$ to the initial hypersurface $S$ meets $S$. Let 
$\phi :G\times S\to S$ denote the action of $G$ on $S$. For $g\in G$ and 
$q\in M$, let $\tilde\phi_g(q)$ be the point which lies the same distance to 
the future of $\phi_g(p)$ along the future-directed geodesic starting
orthogonal to $S$ at $\phi_g(p)$ as $q$ lies to the future 
of $p$ along $\gamma (p)$. This defines a mapping 
$\tilde\phi :G\times M\to M$ by $\tilde\phi (g,q)=\tilde\phi_g(q)$. This
mapping $\tilde\phi$ is an action of $G$ on $M$ which restricts to $\phi$.
By the uniqueness part of Theorem 1.1, it must preserve $g_{\alpha\beta}$ and 
$f$.    

The global theorems to be proved later make use of the concept of
maximal hypersurfaces. A spacelike hypersurface in a spacetime is
called maximal if its mean curvature $\tr k$ is zero. If an initial
data set is given which is maximal in this sense and asymptotically flat
it is of interest to know whether the corresponding local solution of the 
Einstein-Vlasov system, whose existence is guaranteed by the above theorem, 
can be foliated by maximal hypersurfaces in a neighbourhood of the initial
hypersurface. Given what has already been said about the local existence
of asymptotically flat spacetimes, this can be proved using the implicit
function theorem (cf. [16]). The time coordinate $t$ above can be chosen so 
that its level hypersurfaces are maximal hypersurfaces and it can also be 
arranged that as $|x|\to\infty$ the coordinate $t$ agrees asymptotically with 
proper time along a geodesic which starts normal to the initial hypersurface.
(This implies that $g_{00}$ tends to unity as $|x|$ tends to infinity.) When 
this has been imposed the foliation by maximal hypersurfaces is unique and so 
this construction gives a unique preferred time coordinate. The restriction 
of the solution to any of the maximal hypersurfaces is asymptotically flat. 
Once again there is a statement of uniformity. For a family of maximal initial 
data which is bounded in a suitable weighted Sobolev space the maximal 
foliation can be assumed to exist on a time interval which is uniform for 
all data in the family. These statements about the existence of maximal
foliations are not too dependent on the fact that the matter is described
by the Vlasov equation. The essential property which is needed for the
existence and uniqueness theorems is the strong energy condition. 

The approach to studying the global structure of asymptotically flat solutions
of the Einstein-Vlasov system presented in these lectures is closely related 
to work which has been done in the spatially compact case [25, 4]. The
main difference is that in a non-flat spacetime satisfying the strong energy
condition with a compact Cauchy hypersurface there exists at most one 
maximal hypersurface [16]. In this case the maximal foliation of the
asymptotically flat case can be replaced by a constant mean curvature 
(CMC) foliation. This means that each leaf of the foliation has constant
mean curvature, while the value of the mean curvature varies monotonically
from one leaf to the next.

\vskip 10pt\noindent
{\bf 2. Spherical symmetry}

Investigating the global properties of general solutions of the 
Einstein-Vlasov system is beyond the scope of existing mathematical 
techniques. For 
comparison, note that the same comment applies to the Vlasov-Maxwell system
(cf. [19] for the most general known results) while general global existence
results have been obtained for the simpler Vlasov-Poisson system ([17], [13],
[21]). When a system of partial differential equations appears 
inaccessible to direct attack, a natural strategy is to study the simpler 
equations obtained by imposing symmetry conditions on the solutions of the
original equations. In the case of asymptotically flat solutions of the
Einstein-Vlasov system, it seems that there are only two possible symmetry 
assumptions: spherical and rotational symmetry. In the latter case, where the
symmetry group is one-dimensional and has fixed points (on the axis of 
rotation), the simplification obtained is not sufficient to bring the problem
within range of present techniques. Thus in the following treatment of
global questions we consider only the spherically symmetric case. Note for
comparison that in the spatially compact case a wider variety of tractable 
symmetry types exists.

A solution $(M,g_{\alpha\beta}, f)$ of the Einstein-Vlasov system is
said to be spherically symmetric if there exists an action of $SO(3)$
on $M$ by isometries whose generic orbits are two-dimensional such that 
the natural lift 
of this action to the mass shell preserves $f$. There is of course an 
analogous definition of a spherically symmetric initial data set. Consider 
now a spherically symmetric asymptotically flat maximal initial data set. 
{}From the last section we know that there exists a corresponding local 
solution of the Einstein-Vlasov system. Moreover, it can be assumed without 
loss of generality that $SO(3)$ acts on this local solution as a symmetry 
group so that it is spherically symmetric. Furthermore, there exists a 
neighbourhood $U$ of the initial hypersurface which can be foliated by 
maximal hypersurfaces whose intrinsic geometry is asymptotically flat and 
this foliation is unique. It follows from the latter fact that each maximal 
hypersurface is invariant under the action of $SO(3)$. In other words, it is 
a union of orbits of the action of $SO(3)$ on spacetime. Let $r$, the area 
radius, be defined by the condition that on any orbit it takes the constant 
value $\sqrt{A/4\pi}$, where $A$ is the area of the given orbit. Consider now 
a fixed spacelike hypersurface $S_0$ which is invariant under the action of
$SO(3)$. A geodesic of the induced metric on $S_0$ which starts orthogonal to 
the orbits remains orthogonal to them. These geodesics will be called radial 
geodesics. Let $\theta,\phi$ be standard spherical coordinates on one of the 
orbits. Extend them to be constant along the radial geodesics. Since radial 
geodesics can never intersect except at the centre this prescription is 
globally well defined. In the coordinates $(r,\theta,\phi)$ the metric 
components $g_{12}$ and $g_{13}$ vanish. The metric intrinsic to the orbits 
takes the standard form $r^2(d\theta^2+\sin^2 \theta d\phi^2)$. Hence, 
provided the gradient of $r$ does not vanish anywhere, the induced metric on 
$S_0$ can be written in the form
$$e^{2\lambda(r)}dr^2+r^2(d\theta^2+\sin^2\theta d\phi^2)\eqno(2.1)$$
This way of defining spatial coordinates was used in [20]. It cannot be used
if the restriction of $r$ to the hypersurfaces of constant time has a 
vanishing gradient somewhere. A point where the gradient vanishes 
corresponds to a minimal surface. Here we use another type of coordinate 
system which does not suffer from this difficulty. First note that 
it is possible to write the metric in the form
$$B^2(x) dx^2+r^2(x)(d\theta^2+\sin^2 \theta d\phi^2)\eqno(2.2)$$
without restriction, for some function $B$. Since there is always a 
neighbourhood of the origin without minimal surfaces it can be assumed
without loss of generality that $r(x)=x$ near the origin. Since $B$ is
smooth when considered as a function on spacetime, it must be a smooth
function of $x^2$. Furthermore, in order that the spacetime be regular
at $x=0$ and not have a conical singularity, $B(0)$ must be equal to unity.
This means in particular that it is possible to write $B(x)=1+x^2 D(x)$, 
where $D(x)$ is a smooth function of $x^2$. A new coordinate $R$ will be 
sought which is a function of $x$ and has the property that, when expressed 
in terms of the coordinate $R$, the metric takes the form:
$$A^2(R)(dR^2+R^2(d\theta^2+\sin^2 \theta d\phi^2))\eqno(2.3)$$
Coordinates of this sort, or more precisely the Cartesian coordinates
corresponding to these polar coordinates, are often known as isotropic
coordinates. Writing down the coordinate transformation shows that $R(x)$ 
is a solution of the equation $dR/dx=R B(x)/r(x)$. This equation has a solution
which is unique up to a constant scaling. To see this, note first that
this is a first order homogeneous
linear ordinary differential equation. Hence it has
a one-parameter family of solutions which can all be got by multiplying
one particular solution by an arbitrary constant. If we know the existence
of the solution near the origin then the existence for all values of $R$ 
follows by the standard existence and uniqueness theorem for ordinary
differential equations. Near the origin the equation can be solved,
giving $R=Cx\exp\int_0^x x'D(x') dx'$. Near infinity, the asymptotic
flatness of the metric implies that $A$ tends to a constant value. The scaling
can be chosen so that $\lim_{R\to\infty}A(R)=1$ and then the solution is 
determined uniquely. Now let $t$ be a time coordinate which is constant on
each leaf of the preferred foliation by maximal hypersurfaces and which
agrees asymptotically with proper time. Introducing a coordinate $R$ as
above on each leaf puts the metric in the form:
$$-\alpha^2(t,R) dt^2+A^2(t,R)[(dR+\beta (t,R) dt)^2+R^2(d\theta^2
+\sin^2 \theta d\phi^2)]\eqno(2.4)$$
As a consequence of the choice of coordinates the functions $A$ and $\alpha$ 
tend to unity as $R\to\infty$ for each fixed $t$ while $\beta\to 0$. 
The smoothness of the spacetime metric together with spherical symmetry 
implies that $\alpha$, $A$, and $R^{-1}\beta$ are smooth functions of $R^2$. 

For a metric of the form (2.4) which satisfies the extra condition that
the hypersurfaces of constant time are maximal hypersurfaces the field
equations and coordinate conditions take the following form:
$$\eqalignno{
&(R^2(A^{1/2})')'=-\f18 A^{5/2}R^2(\f32 K^2+16\pi\rho)&(2.5)      \cr 
&\alpha''+2\alpha'/R+A^{-1}A'\alpha'=\alpha A^2[\f32 K^2+4\pi(\rho
+\tr S)]&(2.6)                                                    \cr 
&K'+3(A^{-1}A'+1/R)K=8\pi Aj&(2.7)                                \cr 
&\beta'-R^{-1}\beta=\f32\alpha K          &(2.8)                  \cr
&\d_t A=-\alpha KA+(\beta A)'             &(2.9)                  \cr
&\d_t K=-A^{-2}\alpha''+A^{-3}A'\alpha'+\alpha[-2A^{-3}A''
+2A^{-4}{A'}^2-2A^{-3}A'/R-8\pi S_R                  \cr
&\qquad+4\pi\tr S-4\pi\rho]+\beta K'&(2.10)}$$
The notation used in these equations will now be explained. A prime 
denotes a derivative with respect to $R$. The quantity $K$ is that
obtained by contracting the second fundamental form of the hypersurface
$t=$const. twice with the unit vector $A^{-1}\d/\d R$ while $S_R$ is 
obtained in the corresponding way from the energy-momentum tensor.
The quantity $\tr S$ is the trace of the spatial part of the energy-momentum
tensor, i.e. if $n^\alpha$ is the unit future-pointing normal vector 
to the hypersurfaces of constant time, then $\tr S=T_{\alpha\beta}
(g^{\alpha\beta}+n^\alpha n^\beta)$. The quantity $j$ is obtained by
contracting $T_{\alpha\beta}$ once with $n^\alpha$ and once with the
vector $A^{-1}\d/\d R$ and $\rho$ is the energy density $T_{\alpha\beta}
n^\alpha n^\beta$. In the standard terminology of the $3+1$-decomposition
of Einstein's equations, $\alpha$ is the lapse function and $\beta$ is
the one non-trivial component of the shift vector. Equation (2.5) is the 
explicit form of the Hamiltonian constraint (1.7) in this class of spacetimes 
with this kind of coordinate condition. The maximal slicing condition is 
expressed by the lapse equation (2.6). The one non-trivial component of the 
momentum constraint (1.8) in these spacetimes is (2.7). Equation (2.8) is a 
consequence of the coordinate condition chosen while (2.9) follows from the 
definition of the second fundamental form. Finally, (2.10) is the one 
non-trivial Einstein evolution equation in this class of spacetimes. This
form of the field equations has been used by Shapiro and Teukolsky for
numerical calculations (see [27]).

What has been shown above implies that given asymptotically flat spherically
symmetric maximal initial data for the Einstein-Vlasov system, there exists
a corresponding local solution and some $T_1>0$ such that this spacetime
can be covered by coordinates which cast it in the form (2.4) and for which
the time coordinate ranges in the interval $(-T_1,T_1)$ and the initial
hypersurface is given by $t=0$. In fact we are only interested in 
evolution to the future and hence only consider the part of spacetime
on the interval $[0,T_1)$. The quantities describing the metric and matter
then satisfy the equations (2.5)-(2.10). The Vlasov equation is of course
also satisfied. It turns out to be useful for some purposes to write it in
Cartesian coordinates. Let $x^a$ be the coordinates $(R\sin\theta\cos\phi,
R\sin\theta\sin\phi, R\cos\theta)$. Define a related orthonormal frame by
$e_i=A^{-1}\d/\d x^i$. Then if the mass shell is coordinatized by 
$(t,x^a,v^i)$ , where $v^i$ denote the components of a vector in the
given orthonormal frame the Vlasov equation takes the explicit form:
$$\eqalign{
&{\d f\over \d t}+\left(\alpha A^{-1}{v\over\sqrt{1+|v|^2}}-\beta{x\over R}
\right)\cdot {\d f\over\d x}+\left[-A^{-1}\alpha'\sqrt{1+|v|^2}
{x\over R}-{1\over 2}\alpha K\left(v-3v_r{x\over R}\right)\right.    \cr
&\qquad\left. -\alpha A^{-2}A'\left(vv_r-|v|^2{x\over R}\right)
{1\over\sqrt{1+|v|^2}}\right]
\cdot{\d f\over\d v}=0}\eqno(2.11)$$
Here a dot denotes the usual inner product in $\R^3$, $|v|=\sqrt{v\cdot v}$
and $v_r=(v\cdot x)/R$.   
For given initial data there exists a solution on an interval $[0,T_1)$
of the equations (2.5)-(2.11) supplemented by the definitions of the matter
quantities. The whole system of equations will be referred to as the 
\lq reduced Einstein-Vlasov system\rq. This solution of the reduced system
is uniquely determined by its restriction to $t=0$, as follows easily from
the general uniqueness theorem for solutions of the Einstein-Vlasov system
and the uniqueness of the coordinate system used. 
As a consequence there exists a greatest value of $T_1$ (finite or infinite,
call it $T_*$) for which a solution of the reduced equations with the given 
initial data exists on the time interval $[0,T_1)$. The interval $[0,T_*)$ 
is called the maximal interval of existence. The global existence question, 
which is the main theme of these lectures, is the question under what 
circumstances $T_*=\infty$.

One possible strategy for proving global existence theorems will now be
outlined. Suppose that in some way it were possible to show for given
initial data that for any corresponding solution on a finite interval
$[0,T_1)$ the metric components, the distribution function and all
their derivatives of all orders with respect to $t$ and $x$ were
bounded. Then global existence for these initial data would
follow. For the metric components, the distribution function and all
their derivatives of all orders would be uniformly continuous. By a
standard theorem on metric spaces they would all extend to continuous
functions on the closed interval $[0,T_1]$. By another standard
theorem, this time of real analysis, the extensions are $C^\infty$ and
each derivative of each extension is equal to the extension of the
corresponding derivative. In this way smooth initial data are defined
on the hypersurface $t=T_1$. Provided these new initial data are 
asymptotically flat, the local existence theorem can be applied again
to show that the original solution has an extension to an interval
$[0,T_2)$ with $T_2>T_1$. Hence $T_1\ne T_*$. But since $T_1$ was an
arbitrary positive number this only leaves the possibility that $T_*
=\infty$ and global existence is proved. In the following a situation
is exhibited where bounds similar to those which are assumed in
this argument can actually be obtained.

To get closer to the situation which has just been described, consider a
solution of the reduced equations defined on some interval $[0,T_1)$. It
will now be shown that many quantities can be bounded by using the 
Einstein-Vlasov system. By the dominant energy condition $\rho\ge 0$.
Hence equation (2.5) shows that $R^2(A^{1/2})'$ is a non-increasing
function of $R$ for each fixed $t$. When
$R=0$ it is zero and hence $R^2(A^{1/2})'\le 0$. It follows that $A'\le 0$.
The boundary condition that $A\to 1$ as $R\to\infty$ then gives $A\ge 1$.
Next, from (2.6), 
$$(R^2A\alpha')'=\alpha A^3R^2[\f32 K^2+4\pi(\rho+\tr S)]\ge 0\eqno(2.12)$$  
The expression on the right hand side of (2.12) is non-negative, as follows
from the dominant energy and non-negative pressures conditions. Since
$R^2A\alpha'$ vanishes for $R=0$ it can be seen that $R^2A\alpha'\ge 0$ 
and $\alpha'\ge 0$. Using the boundary condition that $\alpha\to 1$ as 
$R\to \infty$ gives $\alpha\le 1$.

Next an estimate of Malec and \'O Murchadha [15] will be used. The
expansions of the null geodesics which start normal to the orbits are 
given by
$$\eqalign{
\theta&=2(A^{-2}A'+(AR)^{-1})+K    \cr
\theta'&=2(A^{-2}A'+(AR)^{-1})-K}\eqno(2.13)$$
The area radius is given by $r=AR$. The theorem of [15] states that,
if the dominant energy condition holds, the quantities $r\theta$ and $r\theta'$
are bounded in modulus by two. Adding and subtracting these estimates and
using the explicit expressions in (2.13) gives the inequalities $RA|K|\le 2$
and $|RA^{-1}A'+1|\le 1$. Since $A\ge 1$ it can be concluded from the first
of these inequalities that $|K|\le 2R^{-1}$. The second inequality gives
$|A^{-1}A'|\le 2R^{-1}$. In particular this gives pointwise bounds for 
$K$ and $A^{-1}A'$ away from the centre. Integrating (2.8) gives
$$R^{-1}_2\beta(R_2)-R^{-1}_1\beta(R_1)=\f32 \int_{R_1}^{R_2} (\alpha K/s) ds
\eqno(2.14)$$
Asymptotic flatness implies that $K=O(R^{-2})$ as $R\to\infty$. Hence it
is possible to let $R_2$ tend to infinity in this equation to get an
expression for $R^{-1}\beta(R)$ as an integral from $R$ to $\infty$.
Using the bounds already obtained for $K$ and $\alpha$ shows that
$$|\beta |\le \f32 R\int^\infty_R (2/s^2) ds\le 3\eqno(2.15)$$
and so $\beta$ is bounded. A bound for $\alpha'$ can be obtained by
analysing the inequality $(R^2A\alpha')'\ge 0$, which was already used to
show that $\alpha\le 1$. Integrating this between the radii $R_1$ and $R_2$
with $R_1<R_2$ gives:
$$\alpha'(R_2)\ge (R_1/R_2)^2 (A(R_1)/A(R_2))\alpha'(R_1)\eqno(2.16)$$
Integrating again then gives:
$$\alpha(R_2)\ge \alpha'(R_1)(R_1^2 A(R_1))\int_{R_1}^{R_2}R^{-2}(A(R))^{-1}
dR\eqno(2.17)$$
Now use the facts that $\alpha(R_2)\le 1$ and that $A(R)\le A(R_1)$ for 
$R\ge R_1$ to see that:
$$1\ge \alpha'(R_1) R_1^2\int_{R_1}^{R_2} R^{-2} dR\eqno(2.18)$$
This holds for all $R_2\ge R_1$ and so it is permissible to replace the
upper limit in the integral by infinity. Evaluating the integral gives 
$\alpha'(R)\le R^{-1}$ for all $R>0$. Note that in deriving all these 
estimates, the only properties of the matter fields used were the dominant 
energy condition and the inequality $\rho+\tr S\ge 0$. The latter follows 
from the strong energy condition. Thus all these estimates hold not only for 
the Einstein-Vlasov system, but also for the Einstein equations coupled to any 
matter model which satisfies the dominant and strong energy conditions. This 
includes perfect fluids with reasonable equations of state, matter described 
by the Boltzmann equation, the massless scalar field (or more generally wave 
maps), any of these matter models combined with an electromagnetic field in 
such a way that the total energy-momentum tensor is the sum of the individual 
energy-momentum tensors, and the Yang-Mills equations for any semi-simple 
gauge group.

These estimates give good information on the solution away from the centre
but, except in the case of the estimate for $\beta$, give no control at the
centre.  Pointwise estimates for $\alpha'$, $A'$ and $K$ which also give 
useful information at the centre can be obtained by optimization arguments.
For any fixed radius $R_0$ we have
$$|K(R)|\le 4\pi A^4(0)\|\rho\|_\infty R_0\eqno(2.19)$$
for any $R\le R_0$. Here $\|\ \|_\infty$ denotes the $L^\infty$ norm in space,
i.e. the maximum value of a function on a hypersurface of constant time.
On the other hand, if $R\ge R_0$ then 
$|K(R)|\le 2R_0^{-1}$. Thus for any value of $R$ it is true that
$$|K(R)|\le 4\pi A^4(0)\|\rho\|_\infty R_0+2R_0^{-1}\eqno(2.20)$$
This can be optimized by choosing $R_0$ so that the function of $R_0$
occurring on the right hand side of this last inequality has a critical
point. This occurs when $R_0$ is equal to 
$[2\pi A^4(0)\|\rho\|_\infty]^{-1/2}$.
It follows that $\|K\|_\infty\le CA^2(0)\|\rho\|_\infty^{1/2}$ for some
constant $C$. A similar procedure can be used to estimate $\alpha'$.
$$\eqalign{
\alpha'(R)&=A^{-1}R^{-2}\int_0^R\alpha A^3 s^2(\f32 K^2+4\pi\rho
+4\pi\tr S)ds                  \cr
&\le CA^7(0)\|\rho\|_\infty R_0}\eqno(2.21)
$$
for $R\le R_0$. Combining this with the previous pointwise estimate for
$\alpha'$ gives
$$\alpha'\le C(A^7(0)\|\rho\|_\infty R_0+R_0^{-1})\eqno(2.22)$$
Doing an optimization as above then leads to an estimate of the form
$\alpha'\le C A^{7/2}(0)\|\rho\|_\infty^{1/2}$. 
{}From the equation for $A$:
$$A'(R)=\f14R^{-2}A^{1/2}\int_0^R A^{5/2}s^2 (\f32 K^2+16\pi\rho) ds
\eqno(2.23)$$
Thus
$$|A'|\le CA(0)[\f14 A^6(0)\|\rho\|_\infty R_0+2R_0^{-1}]\eqno(2.24)$$
Optimizing gives $\|A'\|_\infty\le CA^4(0)\|\rho(0)\|^{1/2}_\infty$.
Starting from the same equations we can also bound $R^{-1}K$, $R^{-1}\alpha'$
and $R^{-1}A'$ pointwise in terms of $\|\rho\|_\infty$ and $A(0)$. In
particular $R^{-1}K$ can be bounded by a constant multiple of 
$A^4(0)\|\rho\|$. Equations (2.6) and (2.7) then allow $\alpha''$ and $K'$ to 
be bounded. Solving (2.5) for $A''$ shows that it too can be bounded. Equation
(2.14) and the bounds for $\beta$ and $R^{-1}K$ imply that $R^{-1}\beta$ can 
be bounded by a constant multiple of $A^2(0)\|\rho\|^{1/2}_\infty$. Using 
equation (2.8) then gives a similar bound for $\beta'$.

In the proof of variants of these estimates discussed in the next section, 
the conservation of the total (ADM) mass plays a role and it is convenient
to say something about this conservation law at this point. It is a
quite general property of asymptotically flat spacetimes. It is particularly
easy to see in the situation considered here, where outside a compact set
the spacetime is vacuum and spherically symmetric. Equation (2.7) can be
rearranged to give $(A^3R^3K)'=8\pi R^3A^4j$. In vacuum this integrates to give
$K=K_0(t)R^{-3}A^{-3}$ for some function $K_0(t)$. Putting this in (2.5) and
using the vacuum condition again gives $(R^2(A^{1/2})')'=O(R^{-4})$. This can 
be integrated to give:
$$A(t,R)=(1+A_0(t)R^{-1})^2+O(R^{-4})\eqno(2.25)$$
It follows from (2.8) that $(R^{-1}\beta)'=\f32 R^{-1}\alpha K=O(R^{-4})$.
Hence $\beta=O(R^{-2})$ and $\beta'=O(R^{-3})$. It then follows from (2.9)
that $\d_t A=O(R^{-3})$. Integrating this last equation in time from $0$ 
to $t$ shows that $A(t,R)=A(0,R)+O(R^{-3})$, so that $A_0(t)$ is in fact
independent of $t$. The ADM mass is given by $\lim_{R\to\infty}(-R^2A')$
and so is also time independent. From (2.5) we can calculate that the
ADM mass is equal to
$$m_{ADM}=\f18\int_0^\infty A^{5/2}R^2(\f32 K^2+16\pi\rho)dR\eqno(2.26)$$
In all these estimates only the dominant and strong energy condtions have
been used.

To go beyond the results of the last paragraphs it is necessary to use
the specific nature of the matter model. For the Einstein-Vlasov system 
a continuation criterion will be proved. It is formulated in terms of 
a quantity $P(t)$, which is defined to be the largest momentum of any 
particle at time $t$. In other words 
$$P(t)=\sup\{|v|: f(t,x,v)\ne 0\ {\rm for\ some}\ x\}\eqno(2.27)$$ 
Before stating the continuation criterion it is necessary to take some time 
to discuss the relation between the differentiability of the functions of
$R$ describing the spacetime and the differentiability of the corresponding
objects in spacetime. The simplest case is that of the scalar functions
$\alpha$ and $A$. They are $C^\infty$ in the spacetime sense if and only
if they are $C^\infty$ as functions of $R$ and all the derivatives of odd
order vanish at the origin. This follows from Lemma A1 of the appendix
with $m=0$. There is also a quantitative version of this, which follows
from Lemma A2. Consider next $\beta$. By definition $\beta=\beta^a x_a/R$
and $\beta^a=\beta x^a/R$, where $\beta^a$ is the shift vector. Because of 
spherical symmetry $\beta$ must vanish at the origin. Hence we can apply 
Lemma A3 to $\beta^a$. Lemma A4 gives quantitative results for $\beta^a$. 
Consider next $K$. By definition $A^2K=k_{ab}x^ax^b/R^2$. The 
maximal hypersurface condition and spherical symmetry together imply that 
$k_{ab}$ vanishes at the origin. Hence Lemma A2 can be applied with $m=2$. 
Similar considerations apply to the matter quantities $j$ and $S_R$, whereby 
in the latter case it is necessary to write $S_R=\tilde S_R+\f13\tr S$, with
$\tilde S_R$ being the contribution to $S_R$ of the trace free part of 
$T_{ab}$. Because of spherical symmetry $\tilde S_R$ vanishes at $R=0$. The 
following expressions for some of the quantities occurring in the Vlasov 
equation are also significant:
$$\eqalign{
\beta x_a/R&=\beta_a                                   \cr
\alpha' x_a/R&=\nabla_a \alpha                         \cr
k_{ab}&=-\f12 KA^2(\delta_{ab}-3x_ax_b/R^2)             \cr
A' x_a/R&=\nabla_a A}\eqno(2.28)$$

\noindent
{\bf Theorem 2.1} If a solution of the reduced Einstein-Vlasov system on the
interval $[0,T)$ for some positive real number $T$ is such that $P(t)$ 
and $A(t,0)$ are bounded then the solution extends to an interval $[0,T_1)$ 
with $T_1>T$. In particular, if the maximal interval of existence $[0,T_*)$ 
is finite then either $P(t)$ or $A(t,0)$ is unbounded there.

\vskip 10pt\noindent
{\bf Proof} Suppose that $P(t)$ is bounded on the interval $[0,T)$. Then
the matter quantities $\rho$, $\tr S$ and $S_R$ are bounded there. It has been
shown above that this, together with the boundedness of $A(t,0)$, implies
that the quantities $A$, $A'$, $A''$, $R^{-1}A'$, $\alpha$, $\alpha'$, 
$\alpha''$, $R^{-1}\alpha'$, $K$, $K'$, $R^{-1}K$. $\beta$, $\beta'$,
and $R^{-1}\beta$ are bounded. It remains to show that all higher spacetime 
derivatives of all these quantities are bounded. We have a $C^2$ bound for
$\alpha$ and $A$ and a $C^1$ bound for $K$ and $\beta$ when these are 
considered as functions of $R$. These imply a $C^2$ bound for $A$, $\alpha$ 
and a $C^1$ bound for $\beta^a$ and $k_{ab}$ in the three dimensional sense, 
using the results of the appendix. It follows that a $C^1$ bound for all the
coefficients of the Vlasov equation on the support of $f$ is obtained. 
The equations obtained by differentiating the Vlasov equation with respect to
$x$ and $v$ then give the boundedness of the first derivatives of $f$ with
respect to $x$ and $v$. Using the definition of the energy-momentum tensor
gives a $C^1$ bound for its Cartesian components. The results of the appendix
then imply a $C^1$ bound for $\rho$, $j$ and $\tr S$. 

Higher derivatives can now be bounded inductively. Assume that a solution
of the reduced equations on a given time interval is such that the $C^{k+1}$
norms of $\alpha$ and $A$ and the $C^k$ norms of $K$, $\beta$, $f$,
$\rho$, $j$, $\tr S$ and $S_R$ are bounded and that $A^{-1}$ is
also bounded. Note that it has already been shown that under the 
hypotheses of the theorem this statement holds for $k=1$ and this suffices to 
start the induction. Now consider the case of general $k$. It is convenient
to rewrite some of the reduced equations in the following form:
$$\eqalignno{
(A^{1/2})'(R)&=-\f18 R^{-2}\int_0^R s^2 [A^{5/2}(\f32 K^2+16\pi\rho)](s) ds 
&(2.29)                 \cr
A(R)\alpha'(R)&=R^{-2}\int_0^R s^2 [A^2 (\f32 K^2+4\pi(\rho+\tr S))](s) ds
&(2.30)                 \cr
A^3(R) K(R)&=R^{-3}\int_0^R s^3 [4\pi A^4j](s) ds
&(2.31)                 \cr
\beta'(R)&=\beta'(0)+R\int_0^R s^{-1}[\f32\alpha K](s) ds
&(2.32)}$$
Applying Lemma A5 to (2.29) and (2.30) gives $C^{k+1}$ bounds for $(A^{1/2})'$
and $A\alpha'$. Combining this with the information already available gives
$C^{k+2}$ bounds for $A$ and $\alpha$. In a similar way, (2.31) and Lemma A5
give a $C^{k+1}$ bound for $K$. The quantity $\beta'(0)$ is already known to
be bounded. Hence (2.32) and Lemma A6 imply a $C^{k+1}$ bound for $\beta$.
(In fact it implies a $C^{k+3}$ bound, but that is not relevant here.) Using
the relationships between differentiability of functions of three variables
and functions of one variable discussed above and the results of the appendix, 
it follows that the coefficients of the Vlasov equation are $C^{k+1}$. Hence
the solution of the Vlasov equation is bounded in the $C^{k+1}$ norm. An
immediate consequence is that the Cartesian components of the energy-momentum
tensor are bounded in the $C^{k+1}$ norm. Finally, applying the results of
the appendix again shows that the $C^{k+1}$ norms of $\rho$, $j$ and $\tr S$
are bounded and this completes the inductive step.      

\vskip 10pt\noindent
It was mentioned earlier that in order to prove global existence it would
suffice to bound the derivatives of all orders of all quantities of interest 
with respect to $t$ and $R$. Here only the derivatives with respect to $R$
have been bounded and it turns out to be difficult to bound the time
derivatives of the lapse function directly. Fortunately it is enough, in
the present context, to bound the spatial derivatives, as will now be shown.
Let $t_n$ be a sequence of times with $t_n<T$ for each $n$ and 
$\lim_{n\to\infty}t_n=T$. The initial data induced by the given solution 
on the hypersurfaces $t=t_n$ define a sequence of initial data which are
bounded in the $C^\infty$ topology. In fact they are also bounded in the
topology of a weighted Sobolev space. To prove this it suffices to obtain
some estimates on an exterior region, say that defined by $R\ge1$. 
Equation (2.29) and the conservation of ADM mass shows that $A'$ is
$O(R^{-2})$, uniformly in $t$. Equation (2.31) and the fact that the support 
of the matter is contained in a region of the form $R\le R_0$ for all $t$ in 
the interval $[0,T_1)$ shows that $K=O(R^{-3})$, uniformly in $t$. Using 
(2.25) and the fact that under the given circumstances the $O(R^{-4})$ error 
term there is uniform in $t$ shows that $A-1=O(R^{-1})$, uniformly in $t$. 
Thus $g_{ab}-\delta_{ab}$ is bounded in $H^1_\delta$ and $k_{ab}$ is bounded
in $H^0_{\delta+1}$ for $-3/2<\delta<-1/2$. To apply the more precise version 
of the local existence theorem for asymptotically flat spacetimes it is 
necessary to have a similar statement for weighted Sobolev spaces of higher 
order. This can be proved straightforwardly by induction using the equations 
(2.5) and (2.7). It follows that the solutions of the Einstein-Vlasov system 
corresponding to the data on each of the hypersurfaces of constant $t$ exist 
on some time interval of length $\epsilon$ about the initial time where data 
are given, with $\epsilon$ independent of $n$. Hence the solution extends to 
the interval $[0,T+\epsilon)$.

\vskip 10pt\noindent
With this result in hand, it is easy to show that the first singularity, if
one exists, must occur in the centre.

\noindent
{\bf Theorem 2.2} If a solution of the reduced Einstein-Vlasov system on the
interval $[0,T)$ for some positive real number $T$ is such that it has a 
smooth extension to an open neighbourhood of the point with coordinates
$(T,0)$ then the solution extends to an interval $[0,T_1)$ with $T_1>T$. 
In particular, if the maximal interval of existence $[0,T_*)$ is finite 
then the solution has a singularity at the point $(T_*,0)$

\vskip 10pt\noindent
{\bf Proof} The neighbourhood occurring in the hypotheses of the theorem 
contains all points with $t>T_1$ and $R\le R_1$ for some $T_1<T$ and some
$R_1>0$. Since the solution is smooth for $t< T_1$ it follows that all
unknowns in the reduced system are bounded on the region $R\le R_1$,
$0\le t<T$. The value of $A$ at any point of a hypersurface of constant
time can be bounded by its value at the centre at the given time and so 
$A$ is bounded on the interval $[0,T)$. It was shown earlier that on any
region of the form $R\ge R_1$ the quantities $A^{-1}A'$ $K$ and $\alpha'$ 
are uniformly bounded. Hence under the present assumptions the quantities 
$A'$, $K$ and $\alpha'$ are bounded on the interval $[0,T_1)$. It was also 
shown that $\beta$ is always bounded everywhere. It can be concluded that 
the coefficients in the Vlasov equation are bounded everywhere. Hence $P(t)$ 
is bounded on the interval $[0,T_1)$, and applying Theorem 2.1 completes the 
proof. 

\vskip 10pt\noindent
{\bf Remark} It is clear from the proof that the existence of an extension
could be replaced by the assumption that there exists some $R_1>0$ such that
$\rho$ and $A$ are bounded on the region $R\le R_1$

\vskip 10pt
Theorems 2.1 and 2.2 are analogues of Theorem 3.2 of [20] and Theorem 4.1
of [22] respectively. It is instructive to compare the theorems involving
maximal-isotropic coordinates proved here with those involving Schwarzschild
coordinates proved in [20] and [22]. The continuation criterion of 
Theorem 2.1 appears at first sight weaker than that of [20] since it is
assumed that not only $P(t)$ but also $A(t,0)$ is bounded. However, it is
much easier to pass from Theorem 2.1 to Theorem 2.2 than it is to pass
from the continuation criterion of [20] to the regularity theorem of [22].
Moreover, the passage from Theorem 2.1 to Theorem 2.2 does not involve any
deep analysis of the Vlasov equation, which the proof of Theorem 4.1 of 
[22] does. Thus it is reasonable to hope that the method of proof used
here can more easily be adapted to matter models other than the Vlasov 
equation than the approach using Schwarzschild coordinates. In the next
section it will be seen that the apparently weaker continuation criterion
given by Theorem 2.1 is also good enough to be applied in the proof of a 
global existence theorem for small initial data.

At this point some further remarks on the notion of \lq well-behaved\rq\
matter models are in order. Consider the case of
dust, i.e. a perfect fluid without pressure. In fact (see [24]), smooth
solutions of the Einstein-dust equations can be considered as
distributional solutions of the Einstein-Vlasov system. Dust satisfies 
the dominant energy and non-negative pressures conditions. However it
cannot be expected that an analogue of Theorem 2.2 holds for dust. The
reason is the occurrence of so-called shell-crossing singularities,
which do not occur at the centre. As has been discussed in [24] and 
[23] this is of significance for the formulation of the cosmic
censorship hypothesis and Theorem 2 can be taken as an indication that
the Einstein-Vlasov system is a good starting point for studying the
cosmic censorship hypothesis and has advantages over other, superficially
simpler matter models, such as a perfect fluid. This is one of the main
motivations for investigating the global properties of solutions of 
these differential equations.   

\vskip 10pt\noindent
{\bf 3. Global existence for small data}

In the last section a continuation criterion was given for solutions of
the reduced equations. Now it will be applied to obtain a global existence
theorem in a particular situation, namely that of small data. The notion
of smallness of initial data is defined in the present context in terms of 
three quantities which characterize the size of the data. Let 
$F_0=\|f(0)\|_\infty$, $P_0=P(0)$ and let $R_0$ be the smallest value of $R$ 
such that $f(0,R)$ vanishes for $R>R_0$.

\noindent
{\bf Theorem 3.1} Let $K$ be a fixed positive constant and consider initial
data for the reduced equations with $R_0\le K$ and $P_0\le K$. Then 
there exists an $\epsilon>0$ such that for all data of this type which,
in addition, satisfy $F_0<\epsilon$ the corresponding solution exists
globally in time and the spacetime which it defines is timelike and null
geodesically complete. 

\noindent
{\bf Remarks} 1. The spacetimes of the theorem are also spacelike geodesically
complete but this will not be proved here. It is the completeness of 
timelike and null geodesics which is most interesting physically, since these
represent the wordlines of particles.
\next
2. The statement on geodesic completeness is an important part of the
theorem since a theorem on global existence in some coordinate time does not
necessarily imply any interesting invariant information.

\vskip 10pt\noindent
In fact more detailed information concerning the asymptotic
behaviour of the spacetimes covered by the theorem will be obtained. In
particular, information will be obtained on the decay of the curvature as
$t\to\infty$. A good understanding of the decay properties of the curvature
is also important for the proof of the theorem and for this reason the
curvature components in a Cartesian frame will now be examined in some 
detail. The curvature can be computed using the following relations (which
are independent of symmetry assumptions):
$$\eqalign{
\st R_{abcd}&=R_{abcd}+k_{ac}k_{bd}-k_{ad}k_{bc}             \cr
\st R_{\sigma bcd}n^\sigma&=-\nabla_c k_{ab}+\nabla_b k_{ac}  \cr
\st R_{\sigma a\tau b}n^\sigma n^\tau&=-8\pi [S_{ab}+\f12 (\rho-\tr S)g_{ab}]
+R_{ab}+\tr k k_{ab}-k_{ac}k^c_b}\eqno(3.1)$$
The first of these equations is the Gauss equation, the second the Codazzi
equation and the third the Einstein evolution equation. It turns out that
the frame components of the curvature tensor which are purely spatial are
linear combinations of the quantities $A^{-3}A''$, $A^{-4}{A'}^2$ and $K^2$, 
with coefficients which are homogeneous functions of the Cartesian 
coordinates. The frame components with two indices corresponding to the time 
direction are linear combinations of these quantities and frame components of 
the energy-momentum tensor. The frame components with precisely one index 
corresponding to the time direction need to be calculated explicitly. They
are given by:
$$\hat R_{0ijk}=\f12 R^{-1}(x_j\delta_{ik}-x_k\delta_{ij})(K'+3KR^{-1}+3A^{-1}
A'K)\eqno(3.2)$$
The combination of metric coefficients which occurs is precisely that which
is familiar from the momentum constraint. 

The idea behind the proof is as follows. In flat space free particles which 
start in a compact set spread out linearly with time. This causes the
associated density to decay. In fact it decays uniformly in space like 
$t^{-3}$ as $t\to\infty$. It is reasonable to suppose that a spacetime 
which evolves from \lq small data\rq\ (i.e. data which are \lq close\rq\
to data for flat space) has small curvature, so that the behaviour
of solutions of the Vlasov equation is similar to that in flat space.
Thus, with luck, the density will fall off at the same rate as in flat
space. Conversely, it is this fall-off of the density which ensures decay
of the curvature. The proof which follows makes these intuitive 
considerations precise and quantitative. It is broken up into a number 
of lemmas which are arranged in a way which parallels as closely as possible 
the proof of the analogous theorem in [20].

\noindent
{\bf Lemma 3.1} Consider a spherically symmetric solution of the reduced
equations on a time interval $[0,T)$ with the property that the support 
of the restriction of $T^{\alpha\beta}$ to each hypersurface $t=$const.
is contained in the ball of radius $R_0+t$ about the centre and  $A(t,0)\le 3$ 
and suppose that:
$$\|\rho\|_\infty (1+t)^{2+\delta}\le K_1\eqno(3.3)$$
for some constants $K_1>0$ and $\delta\in (0,1]$. Then there exists a constant 
$C$, only depending on $K_1$, $\delta$ and the restriction of the solution to 
the initial hypersurface, such that:
$$\|A''\|_\infty (1+t)^{2+\delta}+\|A'\|_\infty (1+t)^{1+\delta}
+\|\alpha'\|_\infty (1+t)^{1+\delta}+\|K\|_\infty (1+t)^{1+\delta}\le C
\eqno(3.4)$$
Moreover the $L^\infty$ norm of the frame components of the curvature tensor
can be bounded by a constant times $(1+t)^{-2-\delta}$.

\noindent
{\bf Proof} Equation (2.29) and the conservation of the ADM mass (2.20)
imply that $A'(R)$ can be bounded by an expression of the form $CR_0^{-2}$
for some constant $C$ for all $R>R_0$. The constant depends a priori on a
pointwise bound for $A(t,0)$ but this is taken care of by the assumption on
$A(t,0)$ occurring in the hypotheses of the lemma. An analogous estimate
holds for $\alpha'$, as a consequence of (2.30). In the case of $K$ it is 
necessary, in order to get an estimate which holds inside the matter, to 
cancel one power of $R$ with a power of $s$ in (2.31). Thus once again the 
estimate obtained involves $R_0^{-2}$. In the last section estimates in terms 
of $R_0^{-1}$ were used in an optimization argument to bound $A'$, $\alpha'$ 
and $K$ by $\|\rho\|_\infty^{1/2}$. If they are replaced by the estimates in 
terms of $R_0^{-2}$ just discussed, then the optimization argument allows the 
$L^\infty$ norms of $A'$, $\alpha'$ and $K$ to be bounded by a constant 
multiple of $\|\rho\|_\infty^{2/3}$. This is enough to take care of the 
last three terms in (3.4). It was remarked in the last section that $R^{-1}A'$
can be bounded pointwise by a constant times $\|\rho\|_\infty$ if $A(t,0)$ is
known to be bounded. Putting this into equation (2.5) together with the
information just obtained gives the desired estimate for the first term 
in (3.4). The estimate for the components of the curvature tensor follows
from the expressions for these components given above.

\noindent
{\bf Remark} The estimates obtained by optimization in the proof of this
lemma could have been used in Section 2 instead of the other estimates
obtained by the same method which were actually used there. The reason for 
presenting both types of estimates is that while the estimates in the proof 
of the lemma are stronger in situations where the solutions are small, the 
estimates of Section 2 are stronger where the solutions are large. Thus they
might be important in other contexts.

The following simple lemma is taken directly from [20].

\noindent
{\bf Lemma 3.2} Consider the ordinary differential equation $du/dt=F(t,u)$
for a $C^1$ function $F$ satisfying the inequality $|F(t,u)|\le\eta
(1+t)^{-1-\delta}(1+|u|)$ for some constants $\delta>0$ and $\eta>0$.
Then given any initial datum at $t=0$ the corresponding solution exists on
the whole of $[0,\infty)$ and satisfies the inequality $|u(t)-u(0)|\le
{\eta\over\delta}\exp {\eta\over\delta}(1+|u(0)|)$.

\vskip 10pt\noindent
This lemma is used to control the behaviour of timelike geodesics 
in a spacetime satisfying certain inequalities. Consider a timelike geodesic
$\gamma$ passing through a point with coordinates $(t,x^a)$ and suppose that
it intersects the initial hypersurface $t=0$. Let $\tau$ be proper time
measured along $\gamma$, starting at $t=0$. Let $\{e'_\sigma\}$ be an 
orthonormal frame which is parallelly transported and is such that
$e'_0$ is the tangent vector to $\gamma$. Let $\theta'^\sigma$ be the dual 
coframe.
A Jacobi field along $\gamma$ is the derivative with respect to the parameter
of a one-parameter family of geodesics in which $\gamma$ is embedded. If it
is expressed as a linear combination $Z^s e'_s$ of $e'_1$, $e'_2$ and $e'_3$
then $Z^s$ satisfies the equation (see [12], p.96):
$$d^2 Z^s/d\tau^2=(R^\alpha_{\beta\gamma\delta} \theta'^s_\alpha e'^\beta_0 
e'^\gamma_t e'^\delta_0)Z^t\eqno(3.5)$$ 
It turns out to be crucial for the global existence theorem to estimate the 
Jacobian determinant of the mapping 
$$v^i\to X^a(0,t,x^a,v^i)\eqno(3.6)$$ 
for fixed values of $t$ and $x^a$, where 
$X^a(s,t,x^a,v^i)$ is part of the solution of the characteristic system, as 
discussed in Section 1. Note that, as indicated by the use of the notation 
$v^i$, the characteristic system which is of interest here is that of the 
Vlasov equation written in terms of frame components. This mapping can be
described in words as follows. Follow the geodesic $\gamma$ through $(t,x^a)$
with initial tangent vector $v^\mu e_\mu$ backwards until it meets the
hypersurface $t=0$. The derivative of this mapping takes vectors tangent
to the mass shell at the point with coordinates $v^i$ to vectors tangent
to the initial hypersurface. A vector tangent to $P$ at $e'_0$ can be
identified with a vector in the tangent space to $M$ at $(t,x^a)$ which
is orthogonal to $e'_0$. This can then be expressed as a linear
combination $Y^s e'_s$. The frame vectors $e'_s$ can be fixed uniquely
by requiring that they be obtained from the coordinate vectors $\d/\d v^i$
by means of the Gram-Schmidt process. The Jacobian of the mapping which
transforms from the basis $\d/\d v^i$ to the basis $e'_s$ and that of its
inverse can be bounded in terms of $|v|$. The derivative of the mapping
(3.6) can conveniently be written as a composition of three linear mappings
from $\R^3$ to itself. The first mapping $L_1$ is the one sending the 
components of a vector tangent to the mass shell in the basis 
$\{\d/\d v^a\}$ to the components $Y^s$. Let $L_2(Y^s)$ be the value at
$t=0$ of the solution of (3.5) with the initial data $Z^s=0$ and $dZ^s
/d\tau=Y^s$ at the point $(t,x^a)$. Let $L_3(Z^s)$ be the Cartesian 
components of the vector obtained by projecting the vector $Z^s e'_s$
onto the hypersurface $t=0$ along the vector $e'_0$. The components
of this vector are given explicitly by:
$$W^a=Z^s[e'^a_s-(e'^0_s/e'^0_0)e'^a_0]\eqno(3.7)$$
the derivative of (3.6) at the point of interest is $L_3L_2L_1$. 
 
\noindent
{\bf Lemma 3.3} Consider spherically symmetric solutions of the 
Einstein-Vlasov system on intervals $[0,T)$ satisfying the following
conditions:
$$\left.\eqalign{
&\|f_0\|_\infty\le\epsilon, \cr
&P_0\le K_2,                \cr
&R_0\le K_3,}\right\}\eqno(3.8)$$
the inequality (3.3) and the condition that $A\le 3$. Then if $\epsilon$ and 
$K_1$ are sufficiently small there exists a constant $K_4$, depending only on
$\epsilon$, $K_1$, $K_2$ and $K_3$, such that $\alpha\ge 1/2$ and:
$$\|T_{\alpha\beta}\|_\infty\le K_4(1+t)^{-3}\eqno(3.9)$$

\noindent
{\bf Proof} Lemma 3.1 provides a lot of information on the decay of geometric 
quantities as $t\to\infty$. In particular the estimate (3.4) can be combined
with Lemma 3.2 to show that if $K_1$ is small enough
$$P(t)\le P_0+1\eqno(3.10)$$
It follows that $\alpha'(R)\le C\epsilon (R+R^{-2})$. If $\epsilon$ is small
enough then this can be integrated to show that $\alpha\ge 1-3C\epsilon/2$.
This implies the first statement of the theorem. The inequality (3.10) 
implies a uniform bound for the Cartesian components
of $e'_0$ if the tangent vector to the geodesic $\gamma$ is contained in
the support of the distribution function. (Only geodesics of this kind are
of interest here.) The remaining frame vectors satisfy:
$$e'^\alpha_{s,\beta}e'^\beta_0+\Gamma^\alpha_{\beta\gamma}e'^\beta_s 
e'^\gamma_0=0\eqno(3.11)$$
The spatial components of (3.11) can be written as
$${d\over d\tau}(e'^a_s)=-\Gamma^a_{\beta\gamma}e'^\beta_s e'^\gamma_0
\eqno(3.12)$$
To profit from this it is necessary to have some information about the
relation between proper time and coordinate time along $\gamma$. In
fact
$$d\tau/dt=\alpha (1+|v|^2)^{-1/2}\eqno(3.13)$$
Since $\alpha\le 1$ the relations (3.12) and (3.13) imply that
$$\left|{d\over dt}(e'^a_s)\right|\le C(|e'^a_s|+|e'^0_s|)(1+t)^{-1-\delta}
\eqno(3.14)$$
On the other hand, since $e'_s$ is a unit vector, $|e'^0_s|\le C|e'^a_s|$.
Consider now the vectors $e'_s$ along all geodesics contained in the support
of $f$. We have a uniform bound for the initial data for equation (3.12)
and so (3.14) and Lemma 3.2 imply the global boundedness of $e'^a_s$ and
hence of $e'^0_s$. (Global boundedness means by definition that they can be 
bounded by a constant which is independent of $T$.)

Consider now once again a timelike geodesic $\gamma$ whose tangent vector 
is contained in the support of the distribution function. As a consequence 
of (3.13) there are positive constants $C_1$, $C_2$ such that
$$C_1 t\le\tau\le C_2 t\eqno(3.15)$$
along $\gamma$. We can assume without loss of generality that $C_2\ge 1$
and then
$$(1+t)^{-2-\delta}\le C_2^{2+\delta}(1+\tau)^{-2-\delta}\eqno(3.16)$$
Thus if $K^s_t=R^\alpha_{\beta\gamma\delta}\theta'^s_\alpha e'^\beta_0
e'^\gamma_t e'^\delta_0$, an estimate of the form
$$|K^s_t(\tau)|\le C(1+\tau)^{-2-\delta}\eqno(3.17)$$
holds along $\gamma$. Let $\tau_0$ be the value of $\tau$ at the point 
$(t,x^a)$ and let $Z^s(\tau)$ be the solution of (3.5) with $Z^s(\tau_0)=0$
and $dZ^s/d\tau (\tau_0)=Y^s$. Let
$$E^s(\tau)=Z^s(\tau)-(\tau-\tau_0)Y^s\eqno(3.18)$$
Then by Taylor's theorem with integral remainder
$$E^s(\tau)=\int_{\tau_0}^\tau (\tau_0-\sigma)(\sigma-\tau)K^s_t(\sigma)
Y^t d\sigma+\int_{\tau_0}^\tau (\tau-\sigma)K^s_t(\sigma)
E^t(\sigma)d\sigma\eqno(3.19)$$
The first integral can be estimated as follows:
$$\eqalign{
\left|\int^{\tau_0}_\tau (\tau-\sigma)(\sigma-\tau_0)K^s_t(\sigma)Y^t
d\sigma\right|&\le(\tau_0-\tau)|Y^t|\int^{\tau_0}_\tau C\sigma
(1+\sigma)^{-2-\delta} d\sigma              \cr
&\le C(\tau_0-\tau)|Y^t|}\eqno(3.20)$$
since $\int_0^\infty \sigma (1+\sigma)^{-2-\delta} d\sigma <\infty$.
Combining (3.19) and (3.20) and applying Gronwall's inequality gives
$$|E^s(\tau)|\le C(\tau_0-\tau)|Y^t|\exp\int_\tau^{\tau_0} C(\sigma-\tau)
(1+\sigma)^{-2-\delta} d\sigma\eqno(3.21)$$
Hence 
$$|E^s(\tau)|\le C(\tau_0-\tau)|Y^t|\eqno(3.22)$$
If $K_1$ is chosen small enough then $C<\f12$ and using (3.22) and the
definition of $E^s$ shows that $|\det L_2|\ge C\tau_0^3$. As a 
consequence of (3.15) $\tau_0$ can be replaced by $t$ in this inequality.
The determinant of $L_1$ can be bounded from below uniformly for all
geodesics whose tangent vectors are contained in the support of the
distribution function. The same is true of the determinant of $L_3$,
as can be seen from (3.7). Thus the statement is obtained that
$$\left|\det\left({\d X^a\over\d v^b}\right)(0,t,x^a,\cdot)\right|\ge Ct^3
\eqno(3.23)$$
This has been shown for all $v^a$ in the support of the distribution function.
If we knew that the mapping (3.6) was injective on the support of $f$ then 
we could change variables from $v^i$ to $X^a$ in the definition of the 
energy-momentum tensor. As a consequence of (3.23) and the boundedness of $P$ 
this would give an estimate of the form (3.9), completing the proof of the 
lemma. 

It will now be shown that for $K_1$ sufficiently small this mapping is indeed 
injective. Suppose that, on the contrary, there are two distinct geodesics 
$\gamma_0$ and $\gamma_1$ starting at the point with coordinates $(t,x^a)$
which meet the initial hypersurface at the same point. Let $v_0^i$ and $v_1^i$
be the components of their initial tangent vectors on the mass shell. For
$l\in [0,1]$ let $v_l^i=(l-1)v_0^i+lv_1^i$. Let $\gamma_l$ be the geodesic
with initial tangent vector corresponding to $v_l^i$. Denote the spatial
coordinates of the point of intersection of $\gamma_l$ with the initial
hypersurface $t=0$ by $\xi^a(l)$. Then $\xi^a(0)=\xi^a(1)$ and $\xi^a(l)$
is a closed curve. Let $w^i=v_1^i-v_0^i$. The tangent vector to the curve 
$\xi^a(l)$ is the image under the mapping $L_3L_2L_1$ of $w^i$. It can be 
concluded from (3.23) that the tangent vector to the curve $\xi^a(l)$ can 
never vanish. It is convenient at this point to make a different choice of
the frame $e'_i$. The change is that instead of the basis $\{\d/\d v^i\}$ of 
the tangent space to the mass shell as above, a basis is chosen which consists 
of vectors which are linear combinations of the vectors $\d/\d v^i$ with
constant coefficients with the first vector of the basis being $w^i\d/\d v^i$.
The vector $L_1(w)$ is proportional to $(1,0,0)$. Using the smallness 
assumption on the data shows that the components $e'^a_1(0)$ are close to the 
components $e'^a_1(\tau_0)$. More precisely, for $K_1$ small enough
$$|e'^a_1 (0)-e'^a_1(\tau_0)|\le \eta |e'^a_1(\tau_0)|\eqno(3.24)$$
for any given $\eta>0$. Similarly
$$|e'^a_0 (0)-e'^a_0(\tau_0)|\le \eta |e'^a_0(\tau_0)|\eqno(3.25)$$
In the same sense $L_2L_1(w)$ is close to a vector of the form $(Z^1,0,0)$.
For $Z^s$ is close to $W^s$ and the frame vectors have also only changed
a little between $\tau=\tau_0$ and $\tau=0$. 
The tangent vector to the curve $\xi^a(l)$ is given by the projection of
$L_2L_1(w)$ on the hypersurface $t=0$ along the vector $e'_0(0)$. The explicit
form of this projection is displayed in equation (3.7). It will now be shown
that, for $K_1$ sufficiently small the second term in the square brackets in
(3.7), with $s=1$, can be bounded in modulus by a constant $k<1$ times the 
modulus of the first term. Before proving this statement it will be shown that
it implies the desired result. $Z^s$ is as close as desired to a vector of the
form $(Z^1,0,0)$ and so $W^a$ is as close as desired to the projection of a 
vector proportional to $e'_1(0)$. The estimate for the terms in (3.7) with
$s=1$ implies that the projection of a vector proportional to $e'_1$ is 
contained in a convex cone about $e'^s_1$. Hence $W^a$ is contained in a 
convex cone about a vector proportional to $w^a$. This is inconsistent with
the fact that $\xi^a(l)$ returns to its starting point. Thus it only remains 
to prove the above statement about the relative sizes of the terms in (3.7). 
Now $e'^a_1$ is close to $A^{-1}Bw^a$ for a certain quantity $B$, $e'^0_1$ is 
close to $B\alpha^{-1}(w\cdot v_l)(1+|v_l|^2)^{-1/2}$, $e'^a_0$ is close to
$A^{-1}v_l^a$ and $e'^0_0$ is close to $\alpha^{-1}(1+|v_l|^2)^{1/2}$. Thus 
the two terms are close to $Bw^a$ and $B(w\cdot v_l)(1+|v_l|^2)^{-1}v_l^a$.
Using the Cauchy-Schwarz inequality the modulus of the second expression 
can be bounded by $|B||w||v_l|^2(1+|v_l|^2)^{-1}$ while the modulus
of the first is $|B||w|$. Using the fact that $|v_l|\le P$, so that
$|v_l|^2/(1+|v_l|)^2\le P^2/(1+P^2)\le k<1$ completes the argument.

\vskip 10pt\noindent
{\bf Lemma 3.4} Suppose that initial data for the reduced system satisfy the 
inequalities (3.8) and let $T_0$ be a fixed positive number. Then for fixed 
$K_2$, $K_3$ there exists $\epsilon>0$ such that the solution corresponding 
to this data exists on the interval $[0,T_0]$, and $P(t)\le 2P(0)$, 
$A(t,0)\le 2$ and $\|\rho\|_\infty$ is as small as desired there.

\noindent
{\bf Proof} First a system of integral inequalities for $P(t)$ and 
$Q(t)=A(t,0)$ will be derived. It is elementary that 
$\|\rho\|_\infty \le C\|f\|_\infty(1+P(t))^4$ with a constant $C$ 
independent of the initial data. It was shown in the last section that the 
coefficients in the part of the characteristic system controlling the 
velocities can be bounded by $C\|\rho\|^{1/2}_{\infty}(1+Q(t))^q$,
for some positive integer $q$. Hence
$$P(t)\le P(0)+C\epsilon^{1/2}\int_0^t P(s)(1+P(s))^2 (1+Q(s))^q ds
\eqno(3.26)$$
Equation (2.9) and the estimates obtained above imply that
$$Q(t)\le Q(0)+C\epsilon^{1/2}\int_0^t P(s)(1+P(s))^p (1+Q(s))^q ds
\eqno(3.27)$$
for some positive integer $p$. Let $z_1$ and $z_2$ be the unique solutions
of the system of integral equations obtained by replacing the inequalities 
in (3.26) and (3.27) by equalities and $P$ and $Q$ by $z_1$ and $z_2$ 
respectively, with initial data $z_1(0)=P(0)$ and $z_2(0)=Q(0)$. When
$\epsilon$ is zero the solution of the integral equations is constant. In
particular, it is global in time. Moreover the initial data $z_1(0)$ and 
$z_2(0)$ can be bounded in terms of $K_2$, $K_3$ and $\epsilon$.
It follows that for $\epsilon$ sufficiently small the solution
of these integral equations exists on a time interval $[0,T_1)$ with
$T_1>T_0$ and, by making $\epsilon$ smaller if necessary, it can be
assumed that $z_1(t)\le 2P(0)$ and $z_2(t)\le 2$ for $0\le t\le T_1$. 
Comparing the integral inequalities with the integral equations shows 
that $P(t)\le z_1(t)$ on any interval where both are defined and similarly
$Q(t)\le z_2(t)$. By the above estimates $\|\rho\|_\infty$ will be as
small as desired if $\epsilon$ is chosen sufficiently small. The continuation 
criterion given by Theorem 2.1 implies that the solution can be extended to 
any time interval where the solution $(z_1,z_2)$ of the integral equations 
exists and this completes the proof of the lemma.

\vskip 10pt\noindent
{\bf Theorem 3.2} Let a non-negative $C^\infty$ compactly supported spherically
symmetric maximal initial datum $f_0$ for the Einstein-Vlasov system be
given which satisfies the inequalities (3.8) for some positive constants
$\epsilon$, $K_2$ and $K_3$. Then if $\epsilon$ is small enough the
corresponding solution of the reduced system exists globally in time.
Moreover, for this solution
$$\|T_{\alpha\beta}\|_\infty \le C(1+t)^{-3}\eqno(3.28)$$
the metric coefficients $\alpha^{-1}$ and $A$ are bounded and the
estimates (3.4) hold with $\delta=1$. 

\noindent
{\bf Proof} Let $K_1$ and $\epsilon$ be positive constants which are small 
enough so that the conclusions of Lemma 3.3 hold for some $\delta<1$. Let
$T_1$ be a positive number satisfying $K_4(1+T_1)^{\delta-1}<K_1$. By Lemma
3.4 the constant $\epsilon$ can be chosen so small that the solution exists
on the time interval $[0,T_1)$. Moreover, it can be arranged that on this 
interval the solution is as small as desired. In particular, $\epsilon$ can
be chosen so that (3.3) is satisfied on the interval $[0,T_1)$ and $A\le 2$
there. Consider a fixed initial datum satisfying (3.8). Define $T_*$ to
be the supremum of those positive numbers $T$ such that the solution of the
reduced equations corresponding to the given initial data exists on $[0,T)$,
and $A\le 2$ for the solution on this interval and it 
satifies (3.3) there. Here $T_*=\infty$ is possible. In fact we will show that 
it is the only possibility. For suppose that $T_*<\infty$. The assumptions 
already made ensure that $T_*>T_1$. The definitions of $T_1$ and $T_*$ and 
Lemma 3.3 then show that the continuation criterion is satisfied on $[0,T_*)$ 
and that
$$\|T^{\alpha\beta}(T_*)\|_\infty \le K_1(1+T_*)^{-2-\delta}\eqno(3.29)$$
This means that the solution can be extended to an interval $[0,T_2)$ with
$T_2>T_*$. Also $T_2$ can be chosen so that (3.3) is satisfied there. This
contradicts the definition of $T_*$ and so in fact it must be the case that
$T_*=\infty$. The inequality (3.3) holds on $[0,\infty)$ for some $\delta<1$.
Applying Lemma 3.1 shows that an inequality of this form also holds for 
$\delta=1$. The remaining conclusions of the theorem then follow from Lemma
3.1.

\vskip 10pt\noindent
The statement of Theorem 3.2 includes all the conclusions of Theorem 3.1
except that concerning geodesic completeness. Equation (3.15) shows that
along a timelike geodesic proper time and coordinate time are equivalent.
This, together with global existence in coordinate time shows that timelike
geodesics are complete. Similarly, the information which we have on the 
geometry is more than enough to show that an affine parameter along a null
geodesic is equivalent to coordinate time.

At the end of the last section some remarks were made about the relation 
of the Vlasov equation with dust. For the Einstein-dust system there is 
no reasonable smallness assumption on initial data which ensures geodesic
completeness of the corresponding solutions, as follows from the work of
Christodoulou [5]. The only other matter model for which a global existence
theorem for spherically symmetric asymptotically flat solutions of the 
Einstein-matter equations with small initial data has been proved is the 
massless scalar field [6]. The method of proof used here in the case of
the Vlasov equation can not be applied directly to the case of matter models
with radiation such as the massless scalar field, since there the flat space
fall-off rates for the matter only make the energy-momentum tensor fall-off
like $t^{-2}$ and do not furnish the faster decay rates used in the above. 

\vfil\eject

\vskip 10pt\noindent
{\bf 4. Local existence in general}

The purpose of this section is to present some aspects of the local existence
theorem for the Einstein-Vlasov system, without any symmetry assumptions.
The standard method for proving such theorems proceeds in several steps.
First, some coordinate conditions are imposed, leading to a system of
\lq reduced equations\rq. This is similar to what was done in Section 2 in
the spherically symmetric case. In the general case different coordinate
conditions are used and hence the reduced system is also different. The
second step is to prove a local existence theorem for the reduced equations.
The third step is to establish the connection between the reduced equations
and the full equations. This means showing that if the data for the reduced
system satisfies certain gauge conditions, if the constraints are satisfied
on the initial hypersurface and if the reduced equations are satisfied 
everywhere, then the coordinate condition and the constraints are satisfied
everywhere. This then implies that the solution of the reduced equations is
actually a solution of the Einstein equations. The first and third steps are 
not discussed further here; the reader is referred to [8] for details. 
(The treatment which follows uses the harmonic coordinate condition, which
is that discussed in [8].)

The reduced equations in harmonic coordinates constitute a system of
nonlinear wave equations for the metric coupled to the Vlasov equation
for the distribution function $f$. Solutions of this system have a domain of 
dependence determined by the light cone and so when proving a local existence 
theorem spatial boundary conditions play no role. For this reason it is 
sufficient to consider data of compact support on $\R^3$. The data for the 
reduced system consists of all components of the metric and their first time 
derivatives together with the distribution function. It is assumed that 
$g_{\alpha\beta}-\eta_{\alpha\beta}$, $\d_t g_{\alpha\beta}$ and $f$ have 
compact support on the initial hypersurface. Such data can never satisfy the 
constraints globally on $\R^3$ except in the case of data for flat space but 
this is not a problem, because of the possibility of using the domain of 
dependence to localize. To start with only the case of $C^\infty$
data is treated but it will be shown later that it is easy to extend
the argument to data of finite differentiability.

Before going further some remarks will be made on the various notations used
for derivatives in this section. This is intended to avoid confusion which
might arise from the mixture of notation from differential geometry and
from analysis which occurs. As in previous sections Greek or Roman indices
attached to geometric objects are usually to be thought of as abstract
indices although they may also occasionally denote components in a coordinate
system. No confusion should arise from this dual role. Indices of this kind
are also used to label the coordinate functions themselves and the partial
derivatives with respect to the coordinates, denoted by $\partial$. The
ranges of the indices are as before. On the other hand the indices on the
operator $D$ denoting differentiation are of a different kind. Here use is
made of the multi-index notation which is very effective in handling
expressions containing high order derivatives of functions of several 
variables. These multi-indices always refer to derivatives with respect to 
the spatial coordinates $x^a$. The expression $D^\alpha u$ denotes a 
particular higher order partial derivative of the function $u$ with repect to 
the spatial variables. The order of this derivative is denoted by 
$|\alpha |$.  Roman indices on $D$ are used in connection with norms as in 
equation (1.11). We have $\|D^i u \|=\max_{|\alpha|=i}\|D^\alpha u\|$. This 
notation may be applied to any norm. 

In order to have a suitable framework for proving a local existence theorem
it is necessary to work with function spaces which are well adapted to the
equation being studied. The natural spaces for hyperbolic equations are 
the Sobolev spaces. In the present context, where the asymptotic behaviour
plays no role, it is possible (and convenient) to use Sobolev spaces without
weights. If $u$ is a smooth function on $\R^3$ define:
$$\|u\|_{H^s}=\left[\sum_{i=0}^s \int |D^i u|^2 dx\right]^{1/2}\eqno(4.1)$$
Compare this with the weighted Sobolev norms defined in (1.11). 

The reduced Einstein equations take the form:
$$g^{\gamma\delta}\d_\gamma\d_\delta g_{\alpha\beta}=F_{\alpha\beta}
(g_{\gamma\delta},
\d_\epsilon g_{\gamma\delta})+8\pi [T_{\alpha\beta}-\f12 g^{\gamma\delta}
T_{\gamma\delta}g_{\alpha\beta}]\eqno(4.2)$$
It is convenient (although not essential) to write this in first order
form by introducing the first derivatives of the metric as additional
variables. Let $h_{\alpha\beta\gamma}=\d_\gamma g_{\alpha\beta}$. Then
in terms of the unknowns $g_{\alpha\beta}$ and $h_{\alpha\beta\gamma}$
the equations can be written as follows:
$$\eqalign{
-g^{00}\d_0 h_{\alpha\beta 0}-2g^{0a}\d_a h_{\alpha\beta 0}
&=g^{ab}\d_a h_{\alpha\beta b}+\ldots                         \cr
g^{ab}\d_0 h_{\alpha\beta a}&=g^{ab}\d_a h_{\alpha\beta 0}    \cr
\d_0 g_{\alpha\beta}&=h_{\alpha\beta 0}}\eqno(4.3)$$
The terms which are not written out explicitly are those coming from
the right hand side of (4.2). The first of these can be written as a
function of $g_{\alpha\beta}$ and $h_{\alpha\beta\gamma}$ without using
derivatives. The reason why certain linear combinations of the original
equations have been taken when writing (4.3) is to make contact with
the notion of a symmetric hyperbolic system. Let $u$ be a function which
takes values in an open set $U$ of $\R^k$ for some $k$. A differential
equation for $u$ of the form:
$$A^0(x^\alpha,u)\d_0 u+A^a(x^\alpha,u)\d_a u+B(x^\alpha,u)=0\eqno(4.4)$$
is called symmetric hyperbolic if the matrices $A^0(x^\alpha,u)$ and 
$A^i(x^\alpha,u)$ are symmetric for all $(x^\alpha,u)$ and if 
$A^0(x^\alpha, u)$ is positive definite. Here $A^0$, $A^a$ are functions on 
the product of $U$ with some open subset of $\R^4$ with values in the 
$k\times k$ matrices and $B$ is a function on the same domain with values in 
$\R^k$. If $u=(g_{\alpha\beta}-\eta_{\alpha\beta},h_{\alpha\beta\gamma})$, 
with $\eta_{\alpha\beta}$ denoting the components of the Minkowski metric
in standard coordinates then the equations (4.3) are of this form, for each 
fixed $T_{\alpha\beta}$. Moreover the reduction process, which is not 
described explicitly here, can be (and usually is) done in such a way that
$g_{00}=-1$ and $g_{0a}=0$ on the initial hypersurface. The open set $U$ is 
defined by the condition that $g_{\alpha\beta}$ have Lorentz signature. The 
Vlasov equation will be written in the form (1.1). Note that the Christoffel 
symbols are rational functions of the unknowns $u$. The full system of reduced
equations consists of (4.3), (1.1) (with the Christoffel symbols expressed 
algebraically in terms of $u$) and the definition (1.3) of $T_{\alpha\beta}$. 
The Vlasov equation itself is a symmetric hyperbolic equation of a 
particularly simple type, with only one unknown. This symmetric hyperbolic 
equation is defined on $\R^6$ rather than $\R^3$. The main theorem can now be 
stated:

\noindent
{\bf Theorem 4.1} Let $(u_0,f_0)$ be $C^\infty$ compactly supported initial
data for the reduced Einstein-Vlasov system written in first order form. Then
there exists a $T>0$ and a $C^\infty$ solution $(u,f)$ of the reduced system 
on the interval $[0,T)$ which induces the given initial data. Moreover $f(t)$
has compact support for each fixed $t$.

\noindent
{\bf Remarks} 1. The domains of definition of the functions $u_0$, $f_0$, $u$
and $f$ are $\R^3$, $\R^6$, $\R^3\times [0,T)$ and $\R^6\times [0,T)$
respectively. 

\noindent
2. The solution $(u,f)$ is the unique solution with the properties
stated in the theorem.

\vskip 10pt
In proving this result an existence theorem for solutions of linear symmetric 
hyperbolic systems with $C^\infty$ initial data is assumed (see e.g.
[9], p. 668). In order to prove Theorem 4.1 a certain iteration is defined 
and then it is proved that this iteration converges in an appropriate sense. 
The convergence follows from certain inequalities satisfied by the solutions of
linear symmetric hyperbolic equations. The proof makes use of the following
Moser-type inequalities [14, 1]. Here these inequalities are only needed
in the case of functions which are smooth and compactly supported. The first 
inequality concerns products of functions and says that if $f$ amd $g$ are 
smooth functions of compact support on $\R^n$ then 
$$\|D^s(fg)\|_2\le C (\|D^s f\|_2\|g\|_\infty+\|f\|_\infty
\|D^s g\|_2)\eqno(4.5)$$
Here $\|\ \|_2$ denotes the $L^2$ norm. The second says that under the same 
conditions for any derivative $D^\alpha$ of order $s$
$$\|D^\alpha(fg)-fD^\alpha g\|_2 \le C(\|D^s f\|_2 \|g\|_\infty
+\|Df\|_\infty \|D^{s-1}g\|_2)\eqno(4.6)$$
The last estimate concerns the composition of a smooth function $F$ with a 
function in a Sobolev space. Suppose that $F$ is a smooth function defined 
on an open subset $U$ of $\R^k$ and that $f$ takes values in an open set $V$ 
of $\R^k$ whose closure is compact and contained in $U$. Then for $s\ge 1$:
$$\|D^s(F(f))\|_2\le C\|F\|_{C^s}\|f\|_\infty^{s-1}\|D^s f\|_2
\eqno(4.7)$$
Here the $C^s$ norm of $F$ is taken over $\bar V$. Note that $s=0$ is 
excluded in (4.7). An inequality for the case $s=0$ can be derived as
follows. Choose some fixed $u_0\in U$. Then there exists a smooth 
matrix-valued function $M$ on $U\times U$ such that $F(u)=M(u,u_0)(u-u_0)$
(cf. [11], p. 77). It follows in an elementary way that:
$$\|F(u)\|_2\le \|F(u_0)\|_2+\|M(u,u_0)\|_\infty \|u-u_0\|_2$$
Now, for fixed $u_0$, $\|M(u,u_0)\|_\infty$ can be bounded in terms of
the $C^1$ norm of $F$ on $\bar V$ for $\|u\|$ in any open set $V$ of the type 
introduced above that contains $\|u_0\|$. If in addition $U$ contains the 
origin and $F(0)=0$ then the inequality reduces to 
$\|F(u)\|_2 \le \|F\|_{C^1}\|u\|_2$, which is a rather close analogue of
(4.7). The fundamental tool used in the existence proof for the nonlinear 
equations is the following standard energy estimate for a linear symmetric 
hyperbolic system. 

\noindent
{\bf Lemma 4.1} Let $u$ be a smooth solution of the linear hyperbolic equation
$A^0(x)\d_0 u+A^a(x)\d_a u+B(x)=0$ whose restriction $u(t)$ to each 
hypersurface $t=$const. has compact support. Suppose further that the functions
$A^\alpha-\bar A^\alpha$ have compact support, where $\bar A^\alpha$ are 
constant matrices with $\bar A^0$ positive definite. Then:
$$\eqalign{
&\|u(t)\|_{H^s}\le\|u(0)\|_{H^s}                    \cr
&+C\int_0^t [(\|A^\mu\|_{C^1}+\|\d_t A^0\|_{C^0})\|u\|_{H^s}
+(\|Du\|_\infty +\|\d_t u\|_\infty )\|A^\mu\|_{H^s}+\|B\|_{H^s} ]
(t') dt'}
\eqno(4.8)$$
where the constant $C$ only depends on upper and lower bounds for the
quadratic form defined by $A^0$.

\noindent
{\bf Proof} Applying the derivative $D^\alpha$ to the equation gives:
$$\eqalign{
&A^0\d_0 (D^\alpha u)+A^a\d_a (D^\alpha u)+D^\alpha B      \cr
&=-[D^\alpha(A^0\d_0 u)-A^0D^\alpha(\d_0 u)]-[D^\alpha(A^a\d_a u)-
A^aD^\alpha(\d_a u)]                                       \cr
&=Q^\alpha,\ {\rm say}.}
\eqno(4.9)$$
This can be used to compute the time derivative of the quantity 
$\int\langle A^0 D^\alpha u,D^\alpha u\rangle$. Differentiating under the
integral and substituting in (4.9) gives a sum of terms of which only one
contains derivatives of $u$ of order higher than that of $D^\alpha$.
However this derivative can be eliminated by partial integration:
$$\int\langle -A^a\d_a (D^\alpha u),D^\alpha u\rangle
=\f12\int\langle\d_a A^a D^\alpha u,D^\alpha u\rangle\eqno(4.10)$$
The equation which results is:
$$d/dt(\int\langle A^0 D^\alpha u, D^\alpha u\rangle)=\int\langle
(\d_t A^0+\d_a A^a)D^\alpha u-2D^\alpha B-2Q^\alpha,D^\alpha u\rangle
\eqno(4.11)$$
Note that the inner product defined by $A^0$ is uniformly equivalent to
the standard inner product on $\R^k$. Hence it follows that, if 
$N^\alpha=\sqrt{\langle A^0 D^\alpha u, D^\alpha u\rangle}$, then
$$|d/dt\int(N^\alpha)^2|\le \int [N^\alpha(\|\d_t A^0
+\d_a A^a\|_\infty N^\alpha+
\|D^\alpha B\|_2+\|Q^\alpha\|_2 )]\eqno(4.12)$$
The $L^2$ norm of $Q^\alpha$ can be estimated with the help of the Moser
estimate (4.6). Using this fact and integrating (4.12) with repect to $t$
gives:
$$\eqalign{
&\int(N^\alpha)^2(t)=\int(N^\alpha)^2(0)              \cr
&\qquad+\int_0^t \int \{N^\alpha(t')
[\|\d_\nu A^\mu\|_\infty \|u\|_{H^s}+\|\d_\nu u\|_\infty
\|DA^\mu\|_{H^{s-1}}+\|D^\alpha B\|_2]\} dt'}\eqno(4.13)$$
Formally, $d/dt((N^\alpha)^2)=2N^\alpha d/dt(N^\alpha)$, so that one factor 
$N^\alpha$ can be cancelled in this formula. This formal calculation can be
justified. Adding the inequalities (4.13) for all derivatives $D^\alpha$
of order less than or equal to some fixed $s$ completes the proof of the
lemma.

\vskip 10pt\noindent
{\bf Proof of theorem 4.1} Define $u_0$ and $f_0$ to be the time independent
functions which agree with the initial data on the hypersurface $t=0$. Then
define an iteration recursively as follows. If $u_n$ and $f_n$ have been
defined, substitute these for $u$ and $f$ in all places in the reduced 
Einstein equations (including the energy-momentum tensor) except those where 
derivatives of $u$ occur. Replace these derivatives by the corresponding 
derivatives of $u_{n+1}$. This gives a linear hyperbolic equation for 
$u_{n+1}$ on an interval $[0,T_{n+1})$ which can be solved with the initial 
datum which is to be prescribed for $u$. Here $T_{n+1}$ is the largest time 
such that the metric defined by certain components of $u_n$ is non-degenerate.
Next substitute $u_{n+1}$ into the Vlasov equation and replace $f$ by 
$f_{n+1}$. This gives a linear hyperbolic equation on the interval 
$[0,T_{n+1})$ which can be solved with the initial datum which is to be 
prescribed for $f$. This defines $u_{n+1}$ and $f_{n+1}$ on the interval 
$[0,T_{n+1})$. In this way approximate solutions $(u_n,f_n)$ of
the reduced system are obtained. They all induce the correct initial data
on the initial hypersurface. Let $C^k([0,T],H^s(\R^n))$ denote the Banach
space of $C^k$ functions on the interval $[0,T]$ with values in the Sobolev
space $H^s(\R^n)$. It will be shown that if $s\ge 5$ and $T>0$ is chosen 
sufficiently small then $T_n\ge T$ for all $n$, the sequence $u_n$ is 
bounded in the space $C^0([0,T],H^s(\R^3))$ and the sequence $f_n$ is bounded 
in the space $C^0([0,T],H^s(\R^6))$. The essential point is that for $s\ge 5$ 
we have the Sobolev embedding theorem which says that in $\R^3$ and $\R^6$ 
any $H^s$ function is $C^1$ and there is a constant $C$ such that 
$\|u\|_{C^1}\le C\|u\|_{H^s}$. Hence, if the estimate of Lemma 4.1 is
applied to the equation for $u_{n+1}$, the pointwise norms of $A^a$ and $u$
occurring there can be estimated in terms of the $H^s$ norms of the same 
quantities. To estimate the norm of $\d_t A^0(u_n)$ which occurs, first
apply the chain rule and then substitute for the term $\d_t u_n$ which 
comes up using the equation. Finally, apply the Sobolev embedding theorem
to the result. The $H^s$ norms of the coefficients $A^0(u_n)$, $A^a(u_n)$
and the part of $B(u_n)$ which does not involve the matter quantities can
be bounded by a polynomial in the $H^s$ norms of $u_n$, using the Moser
estimate (4.7), on any interval where $u$ takes values in a compact subset of
$U$. (Note that this part of $B$ maps the origin to itself and so an
estimate is also obtained for the undifferentiated quantity, as in the
discussion following (4.7).) It is also straightforward to show that, in the 
presence of a bound for the maximum momentum $P(t)$ of any particle in the 
support of the distribution function, the $H^s$ norm of the energy-momentum 
tensor can be estimated by a constant depending on the $H^s$ norm of $u$ 
times the $H^s$ norm of $f$. The result of all this is that if 
$U_n(t)=\sup_{k\le n}\|u_k(t)\|_{H^s}$ and 
$F_n(t)=\sup_{k\le n}\|f_k(t)\|_{H^s}$ then inequalities are obtained of the 
form
$$\eqalign{
U_n(t)&\le U_n(0)+\int_0^t G(U_n(t'),F_n(t')) dt'     \cr
F_n(t)&\le F_n(0)+\int_0^t H(U_n(t'),F_n(t')) dt'}\eqno(4.14)$$
where $G$ and $H$ are polynomials with non-negative coefficients. These 
inequalities are obtained under the assumption that the $u_k$ with $k\le n$ 
take values in a compact subset $K$ of $U$ and that the $P_k$ are all bounded 
by some constant $\bar P$. The functions $F$ and $G$ may depend on $K$ and 
$\bar P$. The inequalities are reminiscent of (3.26) and (3.27) and once again
the solutions of the integral inequalities can be compared with the solutions 
of the corresponding integral equations. Choose the number $T$ that the 
solution of the integral equations with initial data 
$(\|u\|_{H^s},\|f\|_{H^s})$ exists on the interval $[0,T]$. For $n\ge 1$ the 
validity of these inequalities for the functions $u_n$ and $f_n$ depends on 
knowing that the $u_k$ take values in a fixed compact set $K$ of $U$ for all 
$k\le n$. It can be proved by induction that these inequalities are valid for 
$T$ sufficiently small. Note first that they obviously hold for $n=0$. Now 
suppose they hold up to some given value $n$ of the index. Then they give a 
bound on $\|\d_t u_n\|_\infty$. This in turn can be used to show that for $T$
small enough (with a smallness condition which does not depend on $n$)
$u_n$ takes values in a fixed compact set $K$ and $P_{n+1}(t)$ is less than
a fixed constant $\bar P$. This completes the inductive step. It follows that 
$U_n$ and $F_n$ are uniformly bounded on the interval $[0,T]$.
 
Next note that if the coefficients $A^\alpha$ of a symmetric hyperbolic 
system are smooth there exist smooth functions $A^\alpha_1$ such that 
$A^\alpha(u)-A^\alpha(u')=A_1^\alpha(u,u')(u-u')$. A similar statement
applies to the function $B(u)$. Hence the difference $u_{n+1}-u_n$ satisfies 
a linear symmetric hyperbolic equation whose coefficients involve
$A_1^\alpha(u_n,u_{n-1})$. The difference $f_{n+1}-f_n$ satisfies a 
similar equation. Using the estimate of Lemma 4.1, the uniform bounds for
$U_n$ and $F_n$ and the fact that the initial data are the same for all 
iterates leads to an estimate of the form:
$$\|u_{n+1}-u_n\|_{H^{s-1}}+\|f_{n+1}-f_n\|_{H^{s-1}}\le
CT(\|u_n-u_{n-1}\|_{H^{s-1}}+\|f_n-f_{n-1}\|_{H^{s-1}})\eqno(4.15)$$
If $T$ is chosen small enough this shows that the sequences $u_n$ and $f_n$ 
are Cauchy sequences in the Banach space $C^0([0,T],H^{s-1})$ and hence
converge to some limits $(u,f)$ in that space. It should be noted that
the analogue of the estimate (4.15) with $H^{s-1}$ replaced by $H^s$ does
not follow from Lemma 4.1. The reason is that the equation satisfied by the
differences of iterates contain first derivatives of the iterates as 
inhomogeneous terms. To get some statements about convergence in $H^s$, a
little functional analysis will be used. This uses once again the fact that
the iteration is bounded in the $H^s$ norm. The Banach-Alaoglu theorem 
([18],p.115) and the fact that $L^\infty ([0,T],H^s)$ is the dual of a Banach 
space implies that there is a subsequence which converges weakly in 
$L^\infty  ([0,T],H^s)$. It can only converge to $(u,f)$. Thus the limiting 
functions $u$ and $f$ are in $L^\infty ([0,T],H^s)$. If $s$ is chosen to be 
at least seven then $u_n$ and $f_n$ and their spacetime derivatives converge 
uniformly and hence the limiting functions satisfy the equations. It must 
still be checked that if the initial data satisfy the condition 
$\d_\gamma g_{\alpha\beta}=h_{\alpha\beta\gamma}$, then the solution also has 
this property. This holds because the equations (4.3) imply the equation 
$\d_0(h_{\alpha\beta\gamma}-\d_\gamma g_{\alpha\beta})=0$. The time of 
existence depends only on the $H^s$ norm and an argument used in the proof of 
Theorem 2.1 shows that as long as the $H^s$ norm, $s\ge 7$, is bounded on some 
interval the solution can be extended to a longer time interval. If, in 
deriving a differential inequality for the solution, we keep the $C^1$ norms 
instead of eliminating them using the Sobolev embedding theorem, we see that 
the reulting inequalities are linear in the $H^s$ norm. What this means is 
the following. If any $H^s$ norm, $s\ge 7$, is bounded then the $C^1$ norm is 
bounded and then the $H^s$ norm is bounded for every $s$. Hence the solution 
can be extended in the space $H^s$ for every $s$. This implies that a solution
corresponding to $C^\infty$ initial data is itself $C^\infty$ as long as it 
exists in $H^7$. The statement concerning the support of $f$ was already
proved in Section 1 and hence the theorem is proved.

\vskip 10pt\noindent
The estimates used in this proof can also be used to prove a local existence
and uniqueness theorem for data of finite differentiability. The idea is to
approximate the data of finite differentiability by a sequence of $C^\infty$
data which converge to the original data in some Sobolev space $H^s$. By 
Theorem 4.1 there is a $C^\infty$ solution corresponding to each of these 
initial data sets and it can be assumed without loss of generality that all 
these solutions exist on a common time interval $[0,T]$, since the sequence
of data is bounded in $H^s$. If $u$ and $u'$ are solutions of a symmetric 
hyperbolic system then their difference satisfies a homogeneous linear 
symmetric hyperbolic system. The argument is the same as that used to 
estimate the difference of iterates above. The estimate of Lemma 4.1 can
be applied to this equation to show that if $u$ and $u'$ are bounded in
$H^s$ it is possible to bound the $H^{s-1}$ norm of $u-u'$ by a constant
multiple of the corresponding norm of the initial data. A similar estimate
can be obtained for the reduced Einstein-Vlasov system.
It follows that the sequence of $C^\infty$ solutions introduced above
converges to a solution corresponding to the initial data of finite 
differentiability. It may be remarked in passing that the uniqueness of the
solution whose existence is asserted by Theorem 4.1 can also be proved by this
method.

It is possible to prove energy estimates in weighted Sobolev spaces
with a weight $\delta>-3/2$ in a way which is similar to that which
has been done above for ordinary Sobolev spaces. The weighted Moser
estimates which are necessary can be obtained using the techniques of
Bartnik[2] from the Moser estimates for a bounded region in
$\R^n$. These estimates imply the boundedness of a sequence of
approximating solutions in $L^\infty([0,T], H^s_{\delta})$.  Applying
the Banach-Alaoglu theorem once more shows that the previously
obtained solution is in this space if the data are in
$H^s_\delta$. This is still not the most precise result on the
propagation of aymptotic flatness which is desirable. For it is
usually assumed that in initial data sets for the Einstein equations,
the second fundamental form falls off faster than the spatial metric
and it is desirable that this property be preserved by the time
evolution. For a general symmetric hyperbolic system, there is no
obvious reason why this should be true. For if one tries to use the
equation directly to prove a property of this kind, the term $B(u)$
intervenes and in general there is no reason that this should fall off
faster than $u$ itself. However if $B$ at least quadratic it is 
true. Here this is only sketched briefly and we do not even 
give a precise definition of the phrase \lq at least quadratic\rq. 
If $B$ is at least quadratic and $u\in H^s_\delta$, and $\delta>-3/2$,
then $B(u)$ belongs to $H^s_{\delta'}$ for some $\delta'>\delta$. If
$\delta'<\delta+1$ then it follows from the equation that $\d_t u$ is in
$H^{s-1}_{\delta'}$. If $\delta'\ge\delta+1$ then $\d_t u$ is in
$H^{s-1}_{\delta+1}$. If necessary this procedure can be repeated and after
finitely many steps the result is obtained that 
$\d_t u\in H^{s-1}_{\delta+1}$. In the case of the Einstein-Vlasov system,
the matter term in the Einstein equations is irrelevant for this discussion
since its support is in a known compact set. On the other hand, the part of
$B$ only involving the geometry is quadratic in the Christoffel symbols. 
The energy estimates show that the time of existence of the solution in 
a weighted Sobolev space only depends on a weighted Sobolev norm of the
initial data.

\vskip 1cm\noindent
{\bf Appendix}

\noindent
{\bf Lemma A1} Let $f:[0,\infty)\to \R$ be a function, $q$ a polynomial on
$\R^n$ which does not vanish identically and is homogeneous of degree $2m$,
where $m$ is an integer, and let $F$ be the function on $\R^n$ defined by 
$F(x)=f(|x|)q(x)/|x|^{2m}$. If $m\ge 1$ suppose further that $f$ is $C^{2m-2}$
and $f^{(2l)}(0)=0$ for all integers $l$ with $l\le m$. Then for any $k\ge 0$ 
the following conditions are equivalent:

\noindent
(i) $f$ is $C^k$ and $f^{(2l+1)}(0)=0$ for all integers $l$ such that 
$2l+1\le k$
\next
(ii) $F$ is $C^k$

\noindent
{\bf Proof} If (ii) holds consider $g(x^1)=F(x^1,0,\ldots,0)=f(|x^1|)[q(x^1,0,
\dots,0)/(x^1)^{2m}]$. The second factor is a constant. If this constant
is non-zero it can be seen immediately that $f$ is $C^k$. If the constant is
zero it can be made non-zero by a rotation. Since also $g(x^1)=g(-x^1)$, the 
statement (i) holds. Conversely, suppose that (i) holds. Let $p_k$ be the 
Taylor polynomial of order $k$ of $f$ at the origin. Then we can write:
$$F(x)=p_k(|x|)q(x)/|x|^{2m}+(f(|x|)-p_k(|x|))q(x)/|x|^{2m}\eqno({\rm A}1)$$ 
Taylor's theorem applied to $f$ shows that the second term is $C^k$ with
all its derivatives up to order $k$ vanishing at the origin. To complete the 
proof it only remains to show that under the hypothesis (i) the first term is 
a polynomial. Under this hypothesis the first $2m-1$ derivatives of $p_k$ at 
zero vanish, as well as higher derivatives of odd order. Hence $p_k(|x|)$ is 
a polynomial in $|x|^2$ which is divisible by $|x|^{2m}$ and the first term 
in A1 is a polynomial.

\vskip 10pt\noindent 
{\bf Lemma A2} Suppose that the hypotheses of Lemma A1 are satisfied and
that the derivatives of $f$ of order  $\le k$ are bounded. Then the partial
derivatives of $F$ of order $\le k$ are bounded and there exists a constant
$C$ independent of $f$ such that $\|F\|_{C^k}\le \|f\|_{C^k}$. 

\noindent
{\bf Proof} On the region $|x|\ge 1$ the statement is obvious. To
see that it holds when $r\le 1$ is included, first split $F$ into a
sum of two terms as in $A1$ above. The first term and its derivatives
can easily be bounded in terms of derivatives of $f$. A derivative of
the second term of order $\le k$ is a sum of terms, each of which is
the product of a derivative of $(f-p_k)(|x|)$ with a function which is
homogeneous of degree $\ge -k$ and is independent of $f$.  The
derivative of $(f-p_k)(|x|)$ of order $l\le k$ can be bounded on the region 
$|x|\le 1$ in terms of the $\|f\|_{C^l}|x|^l$. The resulting powers of
$|x|$ are sufficient to compensate the functions which are homogeneous
of negative degree and this gives the desired estimate. 

\vskip 10pt\noindent
{\bf Lemma A3} Let $f:[0,\infty)\to \R$ be a function with $f(0)=0$
and let $F^a$ be the vector field on $\R^n$ defined by 
$F^a(x)=f(|x|)x^a/|x|$ for $x\ne 0$ and $F(0)=0$. Then for any 
$k\ge 0$ the following conditions are equivalent:

\noindent
(i) $f$ is $C^k$ and $f^{(2l)}(0)=0$ for all $l$ such that $2l\le k$
\next
(ii) $F^a$ is $C^k$

\noindent
{\bf Proof} $F^1(x^1,0,\dots,0)=f(x^1)$ for $x^1\ge 0$ and this immediately 
shows that (ii) implies that $f$ is $C^k$. Then the relation 
$F^1(-x^1)=-F^1(x^1)$ shows that the derivatives of even order vanish, 
proving (i). Conversely, suppose that (i) holds. The vector field $F^a$ can 
be written as a sum of two terms, as in the proof of Lemma A1, and the second 
of these terms can be handled just as in the proof of that lemma. As for the 
first term, it is the product of $x^a/|x|$ with a polynomial in $|x|$ where 
only the coefficients of odd powers are non-zero. Hence it can be written in 
the form $x^aq(|x|^2)$, for some polynomial $q$. 

\vskip 10pt\noindent 
{\bf Lemma A4} Suppose that the hypotheses of Lemma A3 are satisfied and
that the derivatives of $f$ of order  $\le k$ are bounded. Then the partial
derivatives of $F^a$ of order $\le k$ are bounded and there exists a constant
$C$ independent of $f$ such that $\|F^a\|_{C^k}\le \|f\|_{C^k}$. 

\noindent
{\bf Proof} The method used in the proof of Lemma A2 also applies in this 
case.

\vskip 10pt\noindent
{\bf Lemma A5} Let $f:[0,\infty)\to \R$ be a measurable function and define
$$g_{v,w}(r)=r^{-v}\int_0^r s^w f(s) ds$$
If $w>-1$, $v\le w$ and $f$ is $C^k$ then $g_{v,w}$ is $C^{k+1}$. 
Moreover, on any interval of the form $[0,R]$ there exists a constant $C$, 
independent of $f$, such that $\|g_{p,q}\|_{C^{k+1}}\le C\|f\|_{C^k}$.

\noindent
{\bf Proof} Write $f=p_k+(f-p_k)$, where $p_k$ is the Taylor polynomial of
$f$ of order $k$ at the origin. Now $r^{-v}\int_0^r s^w p_k(s) ds$ can be 
bounded by a constant times $\|f\|_{C^k}$ under the given assumptions.
Thus it can be assumed that without loss of generality that $p_k=0$.
Now:
$$d^l g_{v,w}/dx^l=d^l(r^{-v})/dx^l\int_0^r s^w f(s) ds
+\sum h_j d^{j-1}f/dx^{j-1}$$
where $h_j$ is homogeneous of degree $-v+w+j$. Under the assumption
that $p_k=0$ the integral in the first term is bounded by $Cr^{k+w+1}$,
where the constant $C$ only depends on the $C^k$ norm of $f$. Since the
other factor is $O(r^{-v-l})$ this suffices to bound the first term. The
other terms can be bounded using the fact that $d^{j-1}f/dx^{j-1}$ is
bounded by an expression of the form $C|x|^{k-j+1}$. 

\vskip 10pt\noindent
{\bf Lemma A6} Let $f:[0,\infty)\to \R$ be a continuous function with $f(0)=0$
and define
$$h(r)=r\int_0^r s^{-1} f(s) ds$$
If $f$ is $C^k$ for some $k\ge 1$ then $h$ is $C^{k+1}$. Moreover, on any 
interval of the form $[0,R]$ there exists a constant $C$, independent of $f$, 
such that $\|h\|_{C^{k+1}}\le C\|f\|_{C^k}$.

\noindent
{\bf Proof} Since $f(0)=0$ and $f$ is $C^k$ for some $k\ge 1$ it is 
possible to write $f(r)=rf_1(r)$ for some $C^{k-1}$ function $f_1$.
Then $h(r)=r\int_0^r f_1(s) ds$. Clearly $r^{-1}h$ is $C^k$ and its $C^k$
norm can be estimated by the $C^k$ norm of $f$. It follows that $h$ itself
is $C^{k+1}$ and that its $C^{k+1}$ norm can be bounded by the $C^k$ norm
of $f$. 

\vskip 10pt\noindent
{\bf Acknowledgements} Many of the ideas reported here arose in the course
of my collaboration with Gerhard Rein. I am also indebted to Francis Cagnac
for helpful suggestions.

\vskip 10pt\noindent
{\bf References}

\noindent
[1] Alinhac, S., G\'erard, P.: Op\'erateurs pseudo-diff\'erentiels et
th\'eor\`eme de Nash-Moser. InterEditions, Paris, 1991.
\next
[2] Bartnik, R.: The mass of an asymptotically flat manifold. Commun. Pure
Appl. Math. 34, 661-693 (1986).
\next
[3] Binney, J., Tremaine, S.: Galactic dynamics. Princeton University Press, 
Princeton, 1987.
\next
[4] Burnett, G. A., Rendall, A. D.: Existence of maximal hypersurfaces in 
some spherically symmetric spacetimes. Class. Quantum Grav. 13, 111-123 
(1996).
\next 
[5] Christodoulou, D.: Violation of cosmic censorship in the gravitational
collapse of a dust cloud. Commun. Math. Phys. 93, 171-195 (1984).
\next
[6] Christodoulou, D.: The problem of a self-gravitating scalar field.
Commun. Math. Phys. 105, 337-361 (1986).
\next
[7] Choquet-Bruhat, Y.: Probl\`eme de Cauchy pour le syst\`eme 
int\'egro diff\'erentiel 
\break\noindent
d'Einstein-Liouville. Ann. Inst. Fourier 21, 181-201 (1971).
\next
[8] Choquet-Bruhat, Y., York, J.: The Cauchy problem. In: Held, A. (ed.),
General Relativity and Gravitation, Vol. 1. Plenum, New York, 1980.
\next
[9] Courant, R., Hilbert, D.: Methods of Mathematical Physics, Vol. 2.
Wiley, New York, 1989.
\next
[10] Ehlers, J.: Survey of general relativity theory. In: Israel, W. (ed.)
Relativity, Astrophysics and Cosmology. Reidel, Dordrecht, 1973.
\next
[11] Hamilton, R. S.: The inverse function theorem of Nash and Moser. Bull.
Amer. Math. Soc. 7, 65-222 (1982).
\next
[12] Hawking, S. W., Ellis, G. F. R.: The large-scale structure 
of space-time. Cambridge University Press, Cambridge, 1973.
\next
[13] Lions, P.-L., Perthame, B.: Propagation of moments and regularity
for the three-dimensional Vlasov-Poisson system. Invent. Math. 105,
415-430 (1991).
\next
[14] Majda, A.: Compressible fluid flow and systems of conservation laws in
several space variables. Springer, New York, 1984.
\next
[15] Malec, E., \'O Murchadha, N.: Optical scalars and singularity
avoidance in spherical spacetimes. Phys. Rev. D50, 6033-6036 (1994).
\next
[16] Marsden, J. E., Tipler, F. J.: Maximal hypersurfaces and foliations of
constant mean curvature in general relativity. Phys. Rep. 66, 109-139 (1980).
\next
[17] Pfaffelmoser, K: Global classical solutions of the Vlasov-Poisson
system in three dimensions for general initial data. J. Diff. Eq. 95, 
281-303 (1992). 
\next
{18] Reed, M., Simon, B.: Methods of modern mathematical physics. Academic 
Press, New York, 1972.
\next
[19] Rein, G.: Generic global solutions of the relativistic Vlasov-Maxwell
system of plasma physics. Commun. Math. Phys. 135, 41-78 (1990).
\next
[20] Rein, G., Rendall, A. D.: Global existence of solutions of the 
spherically symmetric Vlasov-Einstein system with small initial data. 
Commun. Math. Phys. 150, 561-583 (1992). Erratum: Commun. Math. Phys.
176:475-478 (1996).
\next
[21] Rein, G., Rendall, A. D.: Global existence of classical solutions to 
the Vlasov-Poisson system in a three dimensional, cosmological setting.
Arch. Rat. Mech. Anal. 126, 183-201 (1994).
\next
[22] Rein, G., Rendall, A. D. and Schaeffer, J.: A regularity theorem
for solutions of the spherically symmetric Vlasov-Einstein system. Commun. 
Math. Phys. 168, 467-478 (1995).  
\next
[23] Rendall, A. D.: On the choice of matter model in general relativity. 
In R. d'Inverno
(ed.) Approaches to Numerical Relativity Cambridge University Press,
Cambridge, 1992. 
\next
[24] Rendall, A. D.: Cosmic censorship and the Vlasov equation. Class. 
Quantum Grav. 9, L99-L104 (1992).
\next
[25] Rendall, A. D.: Crushing singularities in spacetimes with spherical, 
plane and hyperbolic symmetry. Class. Quantum Grav. 12, 1517-1533 (1995).
\next
[26] Swanson, D. G.: Plasma Waves. Academic Press, Boston, 1989.
\next
[27] Shapiro, S. L., Teukolsky, S. A.: Relativistic stellar dynamics on the
computer. I. Motivation and numerical method. Astrophys. J. 298, 34-57
(1985). 
\next
[28] Wald, R. M.: General Relativity. Chicago University Press, Chicago,
1984.
\next

\end